\documentclass{aa}    
\usepackage{txfonts}
\usepackage{longtable}  
\usepackage{natbib,twoopt}
\usepackage{graphicx}

\bibpunct[, ]{(}{)}{,}{a}{}{,}

\usepackage{color}

\newcommand{\Teff}{T_{\rm eff}}
\newcommand{\eps}[1]{\log\varepsilon_{\rm #1}}
\newcommand{\kms}{km\,s$^{-1}$}

\newcommand{\kH}{$S_{\!\!\rm H}$}    
\newcommand{\Eexc}{$E_{\rm exc}$}
\newcommand{\paperI}{\citet[][]{2010A&A...524A..58T}}
\newcommand{\paperII}{\citet[][]{2015A&A...583A..67J}}

\begin{document}

\title{The formation of the Milky Way halo and its dwarf satellites, \\ a NLTE-1D abundance analysis. I. Homogeneous set of atmospheric parameters
}

\author{
  L. Mashonkina\inst{1,2} \and
  P. Jablonka\inst{3,4} \and
  Yu. Pakhomov\inst{2} \and
  T. Sitnova\inst{2} \and
  P. North\inst{3}  
}
 
\offprints{L. Mashonkina; \email{lima@inasan.ru}}
\institute{
     Universit\"ats-Sternwarte M\"unchen, Scheinerstr. 1, D-81679 M\"unchen, 
     Germany \\ \email{lyuda@usm.lmu.de}
\and Institute of Astronomy, Russian Academy of Sciences, RU-119017 Moscow, 
     Russia \\ \email{lima@inasan.ru}
\and Laboratoire d' Astrophysique, Ecole Polytechnique F\'ed\'erale de Lausanne (EPFL), Observatoire de Sauverny, CH-1290 Versoix, Switzerland 
\and GEPI, Observatoire de Paris, CNRS, Universit\'e Paris Diderot, F-92125 Meudon Cedex, France
}

\date{Received   /  Accepted }

\abstract
{

We present a homogeneous set of accurate atmospheric parameters for a
complete sample of very and extremely metal-poor stars in the dwarf
spheroidal galaxies (dSphs)  Sculptor, Ursa Minor, Sextans, Fornax, Bo\"otes~I,
Ursa Major~II, and Leo~IV. We also deliver a Milky Way (MW) comparison sample of
giant stars covering the $-4 <$ [Fe/H] $< -1.7$ metallicity
range.

We show that, in the [Fe/H] $\succsim -3.7$ regime, the non-local
thermodynamic equilibrium (NLTE) calculations with non-spectroscopic
effective temperature ($\Teff$) and surface gravity (log~g) based on
the photometric methods and known distance provide consistent
abundances of the \ion{Fe}{i} and \ion{Fe}{ii} lines. This justifies
the \ion{Fe}{i}/\ion{Fe}{ii} ionisation equilibrium method to
determine log~g for the MW halo giants with unknown
distance.

The atmospheric parameters of the dSphs and MW stars were checked with
independent methods. In the [Fe/H] $> -3.5$ regime, the
\ion{Ti}{i}/\ion{Ti}{ii} ionisation equilibrium is fulfilled in the
NLTE calculations. In the log~g - $\Teff$ plane, all the stars sit on
the giant branch of the evolutionary tracks corresponding to [Fe/H] =
$-2$ to $-4$, in line with their metallicities. For some of the most
metal-poor stars of our sample, we hardly achieve consistent NLTE abundances from the two ionisation stages for both iron and titanium. We suggest that
this is a consequence of the uncertainty in the $\Teff$-colour
relation at those metallicities.  The results of these work provide
the base for a detailed abundance analysis presented in a companion
paper.

}
\titlerunning{Very metal-poor stars in the Milky Way satellites: atmospheric parameters}
\authorrunning{Mashonkina et al.}
\maketitle

%
\section{Introduction}\label{Sect:intro}

How do the first stages of star formation proceed in galaxies? Do
galaxies follow a universal path, independant of their final masses?
What is the level of homogeneity of the interstellar medium from which
stars form? How does it evolve? What is the stellar initial mass
function of the first stars?

These questions can essentially only be addressed in depth in the
Local Group. Only there can we analyse individual stars in sufficient
detail to guide our understanding of the physics of star formation,
supernovae feeback, and the early build-up of galaxies.  The
comparison between ultra-faint, classical dwarf spheroidal galaxies
(UFDs, dSphs), and the Milky Way population offers a fantastic
opportunity to probe different galaxy masses, star formation
histories, levels of chemical enrichement. Both nucleosynthetic
processes and galaxy formation models largely benefit from the
diversity of the populations sampled that way.

We are, however, facing two limitations:

a) Heterogeneity in the samples. Since the first high-resolution
spectroscopic study of the very metal-poor (VMP, [Fe/H]\footnote{In
  the classical notation, where [X/H] = $\log(N_{\rm X}/N_{\rm
    H})_{star} - \log(N_{\rm X}/N_{\rm H})_{Sun}$.} $< -2$) stars in
the Draco, Sextans, and Ursa Minor dSphs \citep{2001ApJ...548..592S}
much of the observational efforts were invested to obtain detailed
chemical abundances of stars in the Milky Way satellites. The largest
samples of the VMP and extremely metal-poor (EMP, [Fe/H] $< -3$)
stars, which were observed with a spectral resolving power of $R >$
20\,000, are available in the literature for the classical dSphs in
Sculptor
\citep{2010A&A...524A..58T,2012AJ....144..168K,2015A&A...583A..67J,2015ApJ...802...93S}
and Ursa Minor
\citep{2004PASJ...56.1041S,2010ApJ...719..931C,2012AJ....144..168K,2015MNRAS.449..761U}.
As for the UFDs, the most studied cases are Bo{\"o}tes~I
\citep{2009A&A...508L...1F,Norris2010,Gilmore2013,2014A&A...562A.146I,2016ApJ...826..110F},
Segue~1 \citep{2014ApJ...786...74F}, Coma Berenices and Ursa Major~II
\citep{2010ApJ...708..560F}. However, the total number of new stars in
each individual paper nowhere exceeds 7.  Therefore, it is common to
combine these samples altogether. However, they were gathered with
different spectroscopic setups and analysed in different ways, with
different methods of the determination of atmospheric parameters, different model atmospheres, radiation transfer and line formation codes, and
line atomic data.  This can lead easily to inaccurate conclusions.

b) Heterogeneity arises also, when applying the LTE assumption for determination of the chemical abundances of the stellar samples with various effective temperatures, surface gravities, and metallicities. Individual stars in the dSphs that are accessible to high-resolution spectroscopy are all giants, and line formation, in particular in the metal-poor atmospheres, is subject to the departures from LTE because of low electron number density and low ultra-violet (UV) opacity. 
For each galaxy, the Milky Way or its satellites, the sampled range of metallicity can be
large \citep{2009ARA&A..47..371T}. Similarly, the position of the
stars along the red giant branch (i.e., their effective temperatures and surface gravities) can vary between samples.

In the literature, determinations of atmospheric parameters and chemical abundances based on the non-local thermodynamic equilibrium (NLTE) line formation were reported 
for Milky Way stars spanning a large interval of metallicities \citep{2013A&A...551A..57H,2013MNRAS.429..126R,Bensby2014,2015ApJ...808..148S,lick_paperII}, however, none has yet treated both the Milky Way and the dSph stellar samples.

In this context, our project aims at providing a homogeneous set of
atmospheric parameters and elemental abundances for the VMP and EMP
stars in a set of dSphs as well as for a Milky Way halo comparison
sample. By employing high-resolution spectral observations and treating the NLTE line formation, our desire is to push the accuracy of the abundance analysis
to the point where the trends of the stellar abundance ratios with
metallicity can be robustly discussed.

In the following, we present the determination of accurate atmospheric
parameters: effective temperatures, $\Teff$, surface gravities,
log~g, iron abundances (metallicity, [Fe/H]), and microturbulence
velocities, $\xi_t$. We rely on the photometric methods, when deriving the effective temperatures. The surface gravities are based on the known distance for the dSph stars and establishing the NLTE ionisation equilibrium between \ion{Fe}{i} and \ion{Fe}{ii} for the Milky Way stars. The metallicities and microturbulence velocities were determined from the NLTE calculations for \ion{Fe}{i-ii}. 
A companion paper focuses on the NLTE
abundances of a large set of chemical elements, spanning from Na to Ba,
and the analysis of the galaxy abundance trends.

The paper is structured as follows: Section~\ref{Sect:basics}
describes the stellar sample and the observational
material. Effective temperatures are determined in
Sect.~\ref{Sect:teff}. In Sect.~\ref{Sect:logg}, we demonstrate that
the \ion{Fe}{i}/\ion{Fe}{ii} ionisation equilibrium method is working
in NLTE down to extremely low metallicities and we derive
spectroscopic surface gravities for the Milky Way (MW) giant sample. The stellar
atmosphere parameters are checked with the \ion{Ti}{i}/\ion{Ti}{ii}
ionisation equilibrium and a set of theoretical evolutionary tracks in
Sect.~\ref{Sect:test}. Comparison with the literature is conducted in
Sect.~\ref{Sect:comparisons}. Section\,\ref{Sect:Conclusions}
summarizes our results.

\section{Stellar sample and observational material}\label{Sect:basics}

 Our sample of VMP stars in dSphs has been selected from published datasets
  by requesting:
\begin{enumerate}
\item   The availability of spectra at high spectral resolution 
  (R = $\lambda / \Delta\lambda \ge$ 25\,000).

\item  Good photometry, enabling
the determination of the atmospheric parameters,
 $\Teff$ and log~g, by non-spectroscopic methods.
\end{enumerate}

We selected 36 stars in total in the classical dSphs Sculptor (Scl), Ursa Minor (UMi),
Fornax (Fnx), and Sextans (Sex) and the ultra-faint dwarfs Bo\"otes~I,
Ursa Major II (UMa~II), and Leo~IV (Table\,\ref{Tab:spectra}). This sample
 covers the $-4 \le {\rm [Fe/H]} < -1.5$ metallicity range.
It is assembled from the following papers:

\begin{itemize}
\item Sculptor: \citet{2015A&A...583A..67J,2012AJ....144..168K,2015ApJ...802...93S}, and \citet{2010A&A...524A..58T},
\item Ursa Minor: \citet{2010ApJ...719..931C,2012AJ....144..168K}, and \citet{2015MNRAS.449..761U},
\item Fornax and Sextans: \citet{2010A&A...524A..58T},
\item Bo{\"o}tes~I: \citet{Gilmore2013,Norris2010}, and \citet[][Boo-980]{2016ApJ...826..110F},
\item UMa~II: \citet{2010ApJ...708..560F},
\item Leo~IV-S1: \citet{2010ApJ...716..446S}. 
\end{itemize}

The comparison sample in the Milky Way halo was selected from the literature based on the following criteria.

\begin{enumerate}
\item The MW and dSph stellar samples should have similar temperatures, luminosities,  and  metallicity range: cool giants with $\Teff \le$ 5250~K and [Fe/H] $< -2$.  
\item High spectral resolution (R $>$ 30\,000) observational material should be accessible
\item Photometry in the V, I, J, K bands was available to derive photometric $\Teff$. 
\end{enumerate}

Binaries, variables, carbon-enhanced stars, and Ca-poor stars were ignored.
  
As a result, the MW comparison sample includes 12 stars from \citet[][hereafter,
  CCT13]{Cohen2013}, two stars from \citet{HE2327,HE2252}, and 
nine stars from \citet{2000ApJ...544..302B}. For the latter subsample we used spectra from the VLT2 / UVES\footnote{http://archive.eso.org/wdb/wdb/adp/phase3-main/query} and
CFHT/ESPaDOnS\footnote{http://www.cadc-ccda.hia-iha.nrc-cnrc.gc.ca/en/search/} archives.

The characteristics of the stellar spectra, which were used in this analysis, are summarized in Table\,\ref{Tab:spectra}. Details on the 
observations and the data reduction can be found in the
original papers. We based our study
on the published equivalent widths ($W_{obs}$s) and line profile fitting, where the observed spectra are available, as indicated in Table\,\ref{Tab:spectra}. 

\longtab{  
\begin{longtable}[]{rllll}  
\caption{\label{Tab:spectra} Characteristics of the used observational material.}  \\
\hline\hline \noalign{\smallskip}
N & Telescope / spectrograph, & Spectral range (\AA), & Objects & Method of analysis \\
  & PIDs                    & R, S/N$^a$            &         &                    \\
\noalign{\smallskip} \hline \noalign{\smallskip}
\endfirsthead
\caption{continued.}\\
\hline\hline \noalign{\smallskip}
N & Telescope / spectrograph, & Spectral range (\AA), & Objects & Method of analysis \\
  & PIDs                    & R, S/N$^a$            &         &                    \\
\noalign{\smallskip} \hline \noalign{\smallskip}
\endhead
\hline
\endfoot
\hline
\endlastfoot
1. & VLT2/UVES,    & 3600-6800,        & Sculptor:       & $W_{obs}$(JNM15), \\
   & 087.D-0928A,  & R $\simeq$ 45\,000, & ET0381, 002\_06,    &  synt \\
   & 091.D-0912A   & S/N = 30-45       & 03\_059, 031\_11,   & \\
    &              &                   & 074\_02    & \\
2. & VLT2/UVES,    & 3800-6800,        & Scl07-49, 07-50,   & $W_{obs}$(TJH10), \\
   & 079.B-0672A,  & R $\simeq$ 45\,000, & Fnx05-42,    &  synt \\
   & 081.B-0620A,  & S/N = 30-40       & Sex11-04, 24-72  & \\
   & 281.B-50220A  &                   &               & \\
3. & Magellan/MIKE, & 3460-9410, &    Sculptor:    & $W_{obs}$(SJF15),\\
   & SJF15          & R $\simeq$ 25\,000, S/N $\simeq$ 80, & 11\_1\_4296, 6\_6\_402, & synt \\
   &                & R $\simeq$ 33\,000, S/N = 171       & S1020549  & \\
4. & Keck~I/HIRES, & 3927-8362,  & Scl 1019417  &  $W_{obs}$(KC12) \\
   & KC12          & R $\simeq$ 29\,300, S/N $\simeq$ 105 & & \\
5. & Keck~I/HIRES, & 3810-6700,  & Ursa Minor:     & $W_{obs}$(CH10) \\
   & CH10          & R $\simeq$ 35\,000, S/N $\simeq$ 80 & COS233, JI19, 28104, & \\
   &               &                                     & 33533, 36886, 41065 & \\
6. & Keck~I/HIRES, & 3927-8362,  & UMi 20103  &  $W_{obs}$(KC12) \\
   & KC12          & R $\simeq$ 34\,500, S/N $\simeq$ 90 & & \\
7. & Keck~I/HIRES, & 4400-7500,  & Ursa Minor:     & $W_{obs}$(UCK15) \\
   & UCK15         & S/N $\simeq$ 25, 30, 12 & 718, 396, 446 & \\
8. & VLT2/FLAMES / UVES, & 4800-5750, 5840-6800, & Bo{\"o}tes~I: & $W_{obs}$(GNM13) \\
   & P82.182.B-0372    & R $\simeq$ 47\,000,   & 33, 41, 94, 117 & \\  
   &                   & S/N = 30, 60          & 127, 130 & \\  
9. & VLT2/UVES, & 3300-4520, 4620-5600, 5680-6650 & Boo-1137 & $W_{obs}$(NYG10) \\
   & P383.B-0038 & R $\simeq$ 40\,000, S/N = 70   &         & \\
10. & Magellan/MIKE, & 3500-9000, R $\simeq$ 28\,000, & Bo{\"o}tes~I: & $W_{obs}$(FNG16), \\ 
   & FNG16          & S/N = 25-30                    & 127, 980      &  synt \\               
11. & Keck~I/HIRES, & 4100-7200,          & UMa~II:    & $W_{obs}$(FSG10) \\
   & FSG10         & R $\simeq$ 37\,000, & S1, S2, S3 &  \\
   &               & S/N = 25-30         &            &  \\
12. & Magellan/MIKE, & 3350-5000, 4900-9300, & Leo IV-S1 & $W_{obs}$(SFM10) \\
   & SFM10          & R $\simeq$ 28\,000, 22\,000, &    & \\
   &                & S/N = 25 &    & \\              
13. & VLT2/UVES,        & 3758-4984, 4780-5758, 5834-6809 & HD~2796 & synt \\
   & 076.D-0546(A) & R $\simeq$ 71\,050, S/N = 339, &         &  \\
   &                   & R $\simeq$ 107\,200, S/N = 256 &         &  \\
14. & VLT2/UVES,        & 3758-4982, 4780-5757, 5834-6808 & HD~8724 & synt \\
   & 71.B-0529(A)  & R $\simeq$ 40\,970, S/N $> 200$ & HD~128279 & \\
   &                   & R $\simeq$ 45\,250, S/N $> 250$ &           & \\
15. & VLT2/UVES,        & 4780-5757, 5834-6808 & HD~108317 & synt \\
   & 68.D-0546(A)  & R $\simeq$ 56\,990, S/N = 100 &  HD~218857  & \\  
16. & VLT2/UVES,    & 3400-4510, S/N = 50 & HE2252-4225 & synt \\
   & 170.D-0010,   & 3756-4978, 4785-5745, 5830-6795 & HE2327-5642 &  \\
   & 280.D-5011    & R $\simeq$ 50\,000, S/N = 100 &           & \\
17. & VLT2/UVES,        & 4774-5758, 5827-6809 & CD$-24^\circ$ 1782  & synt \\
   & 165.N-0276(A)     & R $\simeq$ 80\,930, S/N = 138 &           & \\     
18. & Magellan / MIKE, & 3320-9000, R $\simeq$ 60\,000 & HD~108317 & synt \\
    & Rana Ezzeddine &                               & CD $-24^\circ$ 1782 & \\
19. & CFHT/ESPaDOnS, & 3696-10483 & HD~4306  & synt \\
   & 12BS04         & R $\simeq$ 85\,400, S/N $> 110$ & BD$-11^\circ$ 0145 & \\
20. & VLT2/UVES,    & 3040-10400 & HD~122563   & synt \\
   & UVESPOP       & R $\simeq$ 88\,000, S/N = 460 &           & \\
21. & CFHT/ESPaDOnS, & 3695-10481 & HD~122563 & synt \\
   & 05AC23 (3 spectra) & R $\simeq$ 64\,340, S/N = 93 &    & \\
22. & Keck~I/HIRES, & 3805-5325,          & HE0011-0035 & $W_{obs}$(CCT13) \\
   & CCT13         & R $\simeq$ 36\,000, & HE0332-1007 &  \\
   &               & S/N $> 100$         & HE1356-0622 &  \\
   &               &                     & HE1357-0123 & \\
   &               &                     & HE2249-1704 &  \\
23. & Keck~I/HIRES, & 3250-5990,          & HE0048-0611 & $W_{obs}$(CCT13) \\
   & CCT13         & R $\simeq$ 36\,000, & HE0122-1616 &  \\     
   &               & S/N $\ge 100$       & HE0445-2339 &   \\
   &               &                     & HE1416-1032 &   \\
   &               &                     & HE2244-2116 &   \\
   &              &                     & BS16550-087 &    \\
24. & Magellan / MIKE, & 3350-5000, R $\simeq$ 42\,000, & HE0039-4154 & $W_{obs}$(CCT13) \\ 
   & CCT13          & 4900-9300, R $\simeq$ 32\,000, & HE1416-1032 & \\
   &                & S/N = 120, 90 & & \\
\noalign{\smallskip}\hline \noalign{\smallskip}
\multicolumn{5}{l}{{\bf Notes.} $^a$ The signal-to-noise ratio, as given in the original papers for a wavelength of about 5300\,\AA.} \\ 
\multicolumn{5}{l}{Average S/N is indicated for the MW stars observed with VLT2/UVES and CFHT/ESPaDOnS.} \\
\multicolumn{5}{l}{{\bf Ref.} CCT13 = \citet{Cohen2013}, CH10= \citet{2010ApJ...719..931C}, FNG16 = \citet{2016ApJ...826..110F},} \\
\multicolumn{5}{l}{FSG10 = \citet{2010ApJ...708..560F}, GNM13 = \citet{Gilmore2013}, JNM15 = \citet{2015A&A...583A..67J}, } \\
\multicolumn{5}{l}{KC12 = \citet{2012AJ....144..168K}, NYG10 = \citet{Norris2010}, 
SFM10 = \citet{2010ApJ...716..446S}, } \\
\multicolumn{5}{l}{SJF15 = \citet{2015ApJ...802...93S}, TJH10 = \citet{2010A&A...524A..58T}, UCK15 = \citet{2015MNRAS.449..761U}, } \\
\multicolumn{5}{l}{UVESPOP = \citet{2003Msngr.114...10B}} \\
\end{longtable}
}

\section{Effective temperatures}\label{Sect:teff}

This study is based on photometric effective temperatures. 
We could adopt the published data for about half of our sample, namely:

\begin{itemize}
\item  stars in the Sculptor, Fornax, and Sextans dSphs, for which $\Teff$ was determined in \paperI\ and \paperII\ from the V-I, V-J, and V-K colours and the calibration of \citet{2005ApJ...626..465R} using the CaT metallicity estimates; 
\item stars in Bo\"otes~I, for which $\Teff$ was based on the B-V colour and $griz$ photometry \citep{Norris2010,Gilmore2013};
\item  the CCT13 stellar subsample, with $\Teff$ based on V-I, V-J, and V-K colours that were matched using the predicted colour grid of \citet{2000AJ....119.1448H}; 
\item  HD~122563, for which $\Teff$ is based on angular diameter measurements \citep{2012A&A...545A..17C}.
\end{itemize}

For the rest of the sample, we determined the photometric effective temperatures
ourselves.  The J, H, K magnitudes were taken from \citet[][2MASS All Sky
  Survey]{1538-3881-131-2-1163}, unless an other source is indicated. The
calibration of \citet{2005ApJ...626..465R} was applied and the interstellar
reddening was calculated assuming A$_{\rm V}$ = 3.24 E$_{\rm B-V}$.  The optical
photometry was gathered from a range of sources, as follows.

\begin{itemize}

\item For the star 1019417 in the Sculptor dSph, the star 980 in Bo{\"o}tes~I,
  and the UMa~II stars, we used the
  \citet{2009yCat.2294....0A}'s $ugriz$ magnitudes. They were transformed into V and I
  magnitudes, by applying the empirical colour transformations between the SDSS
  and Johnson-Cousins photometry for metal-poor stars of
  \citet{2006A&A...460..339J}. We checked these transformations on the MW star
  BD~+44$^\circ$2236, for which both the VRI and $gri$ magnitudes are accurate
  within 0.0007~mag and 0.04~mag, respectively. The difference between the
  transformed and observed Johnson-Cousins magnitudes amounts to 0.027~mag,
  hence does not exceed the statistical error given by
  \citet{2006A&A...460..339J}.
  
\item In the Sculptor dSph, for the stars
  11\_1\_4296, 6\_6\_402, and S1020549 we used the V-I and V-K colours and
  metallicities from \citet{2015ApJ...802...93S}. The metallicity of the star 1019417 was taken from \citet{2012AJ....144..168K}.  We adopted E$_{\rm B-V}$ = 0.018 as in \paperI\ and \paperII.

\item In the Ursa Minor dSph, the V and I magnitudes and metallicities were taken from \citet{2010ApJ...719..931C} for the stars COS233, JI19, 28104, 33533, 36886, and 41065, from \citet{2015MNRAS.449..761U} for the stars 396, 446 and 718, and from \citet{2012AJ....144..168K} for the star 20103.
Employing the \citet{1998ApJ...500..525S} maps, we determined a colour excess of E$_{\rm B-V}$ = 0.03. 

\item For the UMa~II dSph stars,
metallicities were taken from \citet{2010ApJ...708..560F}, and E$_{\rm B-V}$ = 0.10 \citep{1998ApJ...500..525S}. 

\item For Boo-980, its effective temperature is based on the V-I colour only, given the large errors of the J and K magnitudes.
  The metallicity is taken from
  \citet{2016ApJ...826..110F}. We adopted a colour excess of E$_{\rm B-V}$ = 0.02
  from \citet{1998ApJ...500..525S}.

\item For Leo~IV-S1, the V-J and V-K colours as well as  E$_{\rm B-V}$ = 0.025 were adopted from \citet[][and private communication]{deJong2010} .

\item For the Milky Way stars HE2252-4225 and HE2327-5642 the photometry was
  taken from \citet{Beersetal:2007} and the metallicities from \citet{HE2252} and
  \citet{HE2327}, respectively. For both stars, a colour excess of E$_{\rm B-V}$ =
  0.013 was adopted according to \citet{1998ApJ...500..525S}.

\item For the remaining eight MW halo giants we used the V magnitudes from the
  VizieR Online Data
  Catalog\footnote{http://vizier.cfa.harvard.edu/viz-bin/VizieR?-source=II/237}
  \citep[][HD~2796, HD~4306, HD~128279]{2002yCat.2237....0D}, the
  Tycho-2
  catalogue\footnote{http://dc.g-vo.org/arigfh/katkat/byhdw/qp/153}
  \citep[][HD~108317, BD~$-11^\circ$ 0145]{2000A&A...355L..27H},
  \citet[][HD~218857]{1985ApJS...58..463N},
  \citet[][HD~8724]{2010A&A...515A.111S}, and \citet[][CD~$-24^\circ$
    1782]{2009A&A...497..497G}.
The colour excess E$_{\rm B-V}$ was estimated for each star from the analysis of
  their position on the (B-V) vs (V-J) diagram. The metallicities were
  taken from \citet{2000ApJ...544..302B}. The final effective temperatures were
  obtained by averaging the individual ones from the V-J, V-H, and V-K
  colours.
\end{itemize}

Table\,\ref{Tab:parameters} lists the adopted effective temperatures.

\longtab{  
\begin{longtable}[]{lcrlcrlccccc}  
 \caption{\label{Tab:parameters} Atmospheric parameters of the selected stars and sources of data.} \\
\hline\hline \noalign{\smallskip}
ID & $\Teff$ & $\sigma_T$ & Method & Ref. & \multicolumn{1}{c}{V} & log~g & $\sigma_{\rm log g}$ & Method & Ref. & [Fe/H]$^1$ &  $\xi_t^1$  \\
   & [K]     & [K]        &        &  & [mag] &       &                      &        &      & & [\kms] \\ 
\noalign{\smallskip} \hline \noalign{\smallskip}
\endfirsthead
\caption{continued.}\\
\hline\hline \noalign{\smallskip}
ID & $\Teff$ & $\sigma_T$ & Method & Ref. & \multicolumn{1}{c}{V} & log~g & $\sigma_{\rm log g}$ & Method & Ref. & [Fe/H]$^1$ &  $\xi_t^1$  \\
   & [K]     & [K]        &        &  & [mag] &       &                      &        &      & & [\kms] \\ 
\noalign{\smallskip} \hline \noalign{\smallskip}
\endhead
\hline
\endfoot
\hline
\endlastfoot
\multicolumn{11}{l}{{\bf Sculptor} classical dSph, \ \ $d$ = 85.9$\pm$4.9~kpc} \\
ET0381      & 4570 &  20 & VIJK & JNM15 & 18.04 & 1.17 & 0.05 & $d^2$ & JNM15 & $-2.19$ & 1.7  \\
002\_06     & 4390 &  70 & VIJK & JNM15 & 17.12 & 0.68 & 0.06 & $d$   & JNM15 & $-3.11$ & 2.3  \\
03\_059     & 4530 &  50 & VIJK & JNM15 & 17.93 & 1.08 & 0.05 & $d$   & JNM15 & $-2.88$ & 1.9  \\
031\_11     & 4670 &  50 & VIJK & JNM15 & 17.80 & 1.13 & 0.05 & $d$   & JNM15 & $-3.69$ & 2.0  \\
074\_02     & 4680 &  70 & VIJK & JNM15 & 18.06 & 1.23 & 0.06 & $d$   & JNM15 & $-3.06$ & 2.0  \\
07-49       & 4630 &  55 & VIJK & TJH10 & 18.35 & 1.28 & 0.05 & $d$   & TJH10 & $-2.99$ & 2.0  \\
07-50       & 4800 & 190 & VIJK & TJH10 & 18.63 & 1.56 & 0.08 & $d$   & TJH10 & $-4.00$ & 2.2  \\
11\_1\_4296 & 4810 & 120 & VIK  & TS    & 19.16 & 1.76 & 0.07 & $d$   & TS    & $-3.70$ & 1.9  \\
6\_6\_402   & 4890 & 170 & VIK  & TS    & 19.13 & 1.78 & 0.08 & $d$   & TS    & $-3.66$ & 1.8  \\
S1020549    & 4650 &  70 & VIK  & TS    & 18.34 & 1.35 & 0.06 & $d$   & TS    & $-3.67$ & 2.0  \\
1019417     & 4280 &  30 & VIJK & TS    & 16.98 & 0.50 & 0.05 & $d$   & TS    & $-2.48$ & 2.0  \\
\multicolumn{11}{l}{{\bf Fornax} classical dSph, \ \ $d$ = 140$\pm$10~kpc} \\
05-42       & 4325 &  70 & VIJHK& TJH10 & 18.48 & 0.70 & 0.07 & $d$   & TJH10 & $-3.37$ & 2.3  \\
 \multicolumn{11}{l}{{\bf Sextans} classical dSph, \ \ $d$ = 90$\pm$10~kpc} \\
11-04       & 4380 & 120 & VIJHK& TJH10 & 17.23 & 0.57 & 0.10 & $d$   & TJH10 & $-2.60$ & 2.2 \\ 
24-72       & 4400 &  40 & VIJHK& TJH10 & 17.35 & 0.76 & 0.09 & $d$   & TJH10 & $-2.84$ & 2.2 \\ 
 \multicolumn{11}{l}{{\bf Ursa Minor} classical dSph, \ \ $d$ = 69$\pm$4~kpc} \\
396         & 4320 &  30 & VIJK &  TS   & 16.94 & 0.70 & 0.05 & $d$   & TS    & $-2.26$ & 2.5 \\ 
446         & 4600 & 220 & VIJK &  TS   & 18.07 & 1.37 & 0.10 & $d$   & TS    & $-2.52$ & 2.5 \\ 
718         & 4630 &  80 & VIJK &  TS   & 17.46 & 1.13 & 0.06 & $d$   & TS    & $-2.00$ & 2.0 \\ 
COS233      & 4370 & 100 & VI   &  TS   & 16.93 & 0.77 & 0.06 & $d$   & TS    & $-2.23$ & 2.0 \\ 
JI19        & 4530 & 100 & VI   &  TS   & 17.26 & 1.00 & 0.06 & $d$   & TS    & $-3.02$ & 2.0 \\ 
20103       & 4780 & 330 & VIJK &  TS   & 18.30 & 1.55 & 0.13 & $d$   & TS    & $-3.09$ & 2.0 \\ 
28104       & 4275 &   5 & VIJK &  TS   & 16.86 & 0.65 & 0.05 & $d$   & TS    & $-2.12$ & 2.0 \\ 
33533       & 4430 & 100 & VI   &  TS   & 16.90 & 0.75 & 0.06 & $d$   & TS    & $-3.14$ & 2.0 \\ 
36886       & 4400 & 100 & VI   &  TS   & 17.01 & 0.82 & 0.06 & $d$   & TS    & $-2.56$ & 2.0 \\ 
41065       & 4350 & 100 & VI   &  TS   & 16.71 & 0.63 & 0.06 & $d$   & TS    & $-2.48$ & 2.0 \\ 
\multicolumn{11}{l}{{\bf Bo{\"o}tes~I} UFD, \ \ $d$ = 60$\pm$6~kpc} \\
 033    & 4730 &  & BV, $griz$ & GNM13 & 17.14$^3$ & 1.4  & & ph$^4$ & GNM13 & $-2.26$ & 2.3  \\
 041    & 4750 &  & BV, $griz$ & GNM13 & 17.34$^3$ & 1.6  & & ph     & GNM13 & $-1.54$ & 2.0  \\
 094    & 4570 &  & BV, $griz$ & GNM13 & 16.25$^3$ & 1.01 & 0.09 & $d$ & TS & $-2.69$ & 2.2  \\
 117    & 4700 &  & BV, $griz$ & GNM13 & 17.10$^3$ & 1.4  & & ph & GNM13 & $-2.09$ & 2.3  \\
 127    & 4670 &  & BV, $griz$ & GNM13 & 17.02$^3$ & 1.4  & & ph & GNM13 & $-1.93$ & 2.3  \\
 130    & 4730 &  & BV, $griz$ & GNM13 & 17.16$^3$ & 1.4  & & ph & GNM13 & $-2.20$ & 2.3  \\
 980    & 4760 &  & VI, $griz$ & TS    & 17.57$^3$ & 1.8  & & NLTE$^5$ & TS  & $-2.94$ & 1.8  \\ 
 1137   & 4700 &  & $griz$     & NYG10 & 17.01$^3$ & 1.39 & 0.09 & $d$ & TS & $-3.76$ & 1.9  \\
 \multicolumn{11}{l}{{\bf UMa~II} UFD, \ \ $d$ = 34.7$\pm$2~kpc} \\
S1          & 4850 & 120 & VIJK & TS    & 17.53 & 2.05 & 0.07 & $d$   & TS    & $-2.96$ & 1.8  \\
S2          & 4780 &  15 & VIJK & TS    & 17.03 & 1.83 & 0.05 & $d$   & TS    & $-2.94$ & 2.0  \\
S3          & 4560 &  15 & VIJK & TS    & 16.02 & 1.34 & 0.05 & $d$   & TS    & $-2.26$ & 1.8  \\
\multicolumn{11}{l}{{\bf Leo~IV} UFD, \ \ $d$ = 154$\pm$5~kpc} \\
S1          & 4530 &  30 & VJK  & TS    & 19.2~ & 1.09 & 0.03 & $d$   & TS    & $-2.58$ & 2.2  \\
\multicolumn{11}{l}{{\bf Milky Way} halo} \\
HD~2796     & 4880 & 46 & VJHK   & TS    & 8.50 & 1.55 & & NLTE & TS    & $-2.32$ & 1.8 \\
HD~4306     & 4960 & 54 & VJHK   & TS    & 9.02 & 2.18 & & NLTE & TS    & $-2.74$ & 1.3 \\
HD~8724     & 4560 & 45 & VJHK   & TS    & 8.34 & 1.29 & & NLTE & TS    & $-1.76$ & 1.5 \\
HD~108317   & 5270 & 48 & VJHK   & TS    & 8.03 & 2.96 & & NLTE & TS    & $-2.18$ & 1.2 \\
HD~122563   & 4600 & 41 & Int$^6$ & CTB12 & 6.19 & 1.32 & & NLTE & TS & $-2.63$ & 1.7  \\
HD~128279   & 5200 & 72 & VJHK   & TS    & 8.00 & 3.00 & & NLTE & TS    & $-2.19$ & 1.1  \\
HD~218857   & 5060 & 46 & VJHK   & TS    & 8.95 & 2.53 & & NLTE & TS    & $-1.92$ & 1.4  \\
HE0011-0035 & 4950 & 25 & VIJK & CCT13 & 15.04 & 2.0 & & NLTE & TS & $-3.04$ & 2.0 \\
HE0039-4154 & 4780 & 42 & VIJK & CCT13 & 13.92 & 1.6 & & NLTE & TS & $-3.26$ & 2.0 \\ 
HE0048-0611 & 5180 & 121& VIJK & CCT13 & 15.47 & 2.7 & & NLTE & TS & $-2.69$ & 1.7  \\
HE0122-1616 & 5200 & 11 & VIJK & CCT13 & 15.77 & 2.65 & & NLTE & TS & $-2.85$ & 1.8  \\
HE0332-1007 & 4750 & 15 & VIJK & CCT13 & 14.59 & 1.5 & & NLTE & TS & $-2.89$ & 2.0  \\
HE0445-2339 & 5165 & 66 & VIJK & CCT13 & 14.08 & 2.2 & & NLTE & TS & $-2.76$ & 1.9  \\
HE1356-0622 & 4945 & 98 & VIJK & CCT13 & 14.31 & 2.0 & & NLTE & TS & $-3.45$ & 2.0  \\
HE1357-0123 & 4600 & 75 & VIJK & CCT13 & 14.74 & 1.2 & & NLTE & TS & $-3.92$ & 2.1  \\
HE1416-1032 & 5000 & 76 & VIJK & CCT13 & 15.03 & 2.0 & & NLTE & TS & $-3.23$ & 2.1  \\
HE2244-2116 & 5230 & 150& VIJK & CCT13 & 15.75 & 2.8 & & NLTE & TS & $-2.40$ & 1.7  \\
HE2249-1704 & 4590 & 33 & VIJK & CCT13 & 15.25 & 1.2 & & NLTE & TS & $-2.94$ & 2.0  \\
HE2252-4225 & 4750 & 80 & VIJK & TS    & 14.88 & 1.55 & & NLTE & TS & $-2.76$ & 1.9  \\
HE2327-5642 & 5050 & 80 & VIJK & TS    & 13.88 & 2.20 & & NLTE & TS & $-2.92$ & 1.7  \\
BD $-11^\circ$ 0145 & 4900 & 72 & VJHK & TS & 10.72 & 1.73 & & NLTE & TS    & $-2.18$ & 1.8 \\
CD $-24^\circ$ 1782 & 5140 & 52 & VJHK & TS & 9.97 & 2.62 & & NLTE & TS    & $-2.72$ & 1.2 \\
BS16550-087 & 4750 & 56 & VIJK & CCT13 & 13.75 & 1.5 & & NLTE & TS & $-3.33$ & 2.0  \\
\noalign{\smallskip}\hline \noalign{\smallskip}
\multicolumn{11}{l}{{\bf Notes.} $^1$ from NLTE analysis of the iron lines, this study; $^2$ V or $i$ and known distance; $^3$ SDSS $i$ magnitude;} \\
\multicolumn{11}{l}{ \ \ $^4$ photometry and the YY Isochrones; $^5$ \ion{Fe}{i}/\ion{Fe}{ii}, NLTE; $^6$ interferometry.} \\
\multicolumn{11}{l}{{\bf Ref.} CCT13 = \citet{Cohen2013}, CTB12 = \citet{2012A&A...545A..17C}, GNM13 = \citet{Gilmore2013}, } \\
\multicolumn{11}{l}{JNM15 = \citet{2015A&A...583A..67J}, NYG10 = \citet{Norris2010}, TJH10 = \citet{2010A&A...524A..58T}, } \\
\multicolumn{11}{l}{TS = this study.} \\
\end{longtable}
}

\section{Surface gravities}\label{Sect:logg}

We need to apply two different methods to determine surface gravities of our stellar sample. The determination of log~g of the dSph stars benefits from their common distance. Most of the Milky Way stars have no accurate distances, and we shall rely
on the spectroscopic method that is based on the NLTE analysis of lines of iron in the two ionisation stages. Using the dSph stars with non-spectroscopic log~g, we prove that the \ion{Fe}{i}/\ion{Fe}{ii} ionisation equilibrium method is working for VMP and EMP giants.

\subsection{Photometric methods}\label{Sect:ph}

Surface gravity of the dSph stars can be calculated by
applying the standard relation between log~g, $\Teff$, the absolute bolometric
magnitude M$_{bol}$, and the stellar mass $M$. This is the method that we rely
on in this study, and we denote such a gravity log~g$_d$. We assumed $M = 0.8
M_\odot$ for our RGB sample stars. The adopted distances are as follows.

\begin{itemize}
  
\item Sculptor, Fornax, and Sextans: $d$ = 85.9~kpc,
140~kpc, and 90~kpc, respectively, taken from  \citet{2015A&A...583A..67J} and
\citet{2010A&A...524A..58T}

\item Leo~IV-S1: $d = 154\pm5$~kpc \citep{2009ApJ...699L.125M}. 

\item Ursa Minor:  $d$ = 69$\pm$4~kpc \citep{1999AJ....118..366M}.

\item UMa~II: $d = 34.7\pm2$~kpc \citep{2012ApJ...752...42D}.

\item Bo\"otes~I: $d$ = 60$\pm$6~kpc \citep{2006ApJ...647L.111B}.
\end{itemize}
 
In case of the Sculptor, Fornax, and Sextans dSphs, we used the log~g$_d$ values derived by \citet{2015A&A...583A..67J} and \citet{2010A&A...524A..58T}, which
were obtained with the bolometric correction of \citet{Alonso1999}. For the Sculptor dSph, \citet{2008AJ....135.1993P}
derived statistical and systematic errors of the distance modulus as 0.02~mag
and 0.12~mag, respectively, leading to a maximum shift of 0.05~dex in log~g$_d$. An
uncertainty of 80~K in $\Teff$ results in uncertainty of 0.03~dex in
log~g$_d$.

The same method was applied to most of the rest of our dSph stars. If we used
the V magnitudes, then we adopted the bolometric correction from
\citet{Alonso1999}. If the SDSS $i$ magnitude was used, then the bolometric
correction was from \citet{2014MNRAS.444..392C}.  The sources of photometry were
cited in the previous section.

Statistical error of the distance based surface gravity was computed as the quadratic sum of errors of the star's distance, effective temperature, mass, visual magnitude, and bolometric correction: 

\begin{equation}
\sigma_{\rm log g}^2 = (2 \sigma_{\log d})^2 + (4 \sigma_{\log T})^2 + \sigma_{\log M}^2 + (0.4 \sigma_V)^2 + (0.4 \sigma_{\rm BC})^2 . 
\end{equation}

\noindent Here, $\sigma_{\log d}$ is taken for given dSph and $\sigma_{\log T}$ for each individual star, while we adopt common uncertainty of 0.02~M$_\odot$ in the star's mass, $\sigma_V$ = 0.02~mag, and $\sigma_{\rm BC}$ = 0.02~mag.

Another method to determine the surface gravity relies on placing stars on
isochrones. The gravities derived in this way are denoted log~g$_{ph}$.

In Bo\"otes~I, log~g$_{ph}$ were determined together with $\Teff$ by \citet{Norris2010}
and \citet[][adopting the NY analysis]{Gilmore2013} from the $(g - r)_0$ and $(r
- z)_0$ colours, assuming that the stars were on the red giant branch and
iteratively using the synthetic $ugriz$ colours of
Castelli\footnote{http://wwwuser.oat.ts.astro.it/castelli/colors/sloan.html} and
the Yale-Yonsei Isochrones
\citep{2004ApJS..155..667D}\footnote{http://www.astro.yale.edu/demarque/yyiso.html},
with an age of 12~Gyr.

For most of the Bo\"otes~I stars, the absolute difference between log~g$_d$ and
log~g$_{ph}$ does not exceed 0.06~dex. Therefore, we adopted the original
surface gravities of \citet{Gilmore2013} as final. In contrast, for Boo-94 and
Boo-1137, we found log~g$_d$ greater than log~g$_{ph}$ by 0.21~dex and 0.19~dex,
respectively. As shown in Sect.\,\ref{sect:sp}, log~g$_d$ leads to consistent
NLTE abundances from lines of \ion{Fe}{i} and \ion{Fe}{ii} in Boo-94 and smaller
difference between the two ionisation stages for Boo-1137. Consequently, we
adopted the log~g$_d$ value as final surface gravity for these two stars.

As to the Milky Way sample, \citet{Cohen2013} applied an approach
similar to that of \citet{Norris2010} and \citet{Gilmore2013} to
determine photometric log~g$_{ph}$ values for their stellar sample, 
using the VIJK photometry.

\subsection{Spectroscopic methods}\label{Sect:Method}

\subsubsection{Line selection and atomic data}

Following \citet{2015A&A...583A..67J}, we did not use the \ion{Fe}{i} lines with
low excitation energy of the lower level, \Eexc\ $<$ 1.2~eV. This is because our study is based on classical plane-parallel (1D) model atmospheres, while the low-excitation levels are predicted to be affected by hydrodynamic phenomena (3D effects) in the atmosphere to more
degree than the higher excitation lines \citep{Collet2007,Hayek2011,2013A&A...559A.102D}.
For example, in
the 4858 / 2.2 / $-3$ model, the abundance correction (3D-1D) amounts to $-0.8$~dex
and $0.0$~dex for the \ion{Fe}{i} lines arising from \Eexc\ = 0 and 4~eV,
respectively \citep[][$W$ = 50\,m\AA]{Collet2007}. We do not see in general such
a large discrepancy between the low- and high-excitation lines of \ion{Fe}{i} in
the investigated stars, nevertheless, for example, in Scl07-50 the difference in
LTE abundances between the \Eexc\ $<$ 1.2~eV and \Eexc\ $>$ 1.2~eV lines amounts
to 0.36~dex.

The spectral lines used in the abundance analysis are listed in
  Table\,\ref{Tab:linelist} (Appendix) together with their atomic
  parameters. The $gf$-values and the van der Waals damping constants, $\Gamma_6$, based on
the perturbation theory \citep{2000A&AS..142..467B} 
  were taken from VALD3 \citep{2015PhyS...90e4005R}, at the exception of \ion{Fe}{ii}, for which we used the
  $gf$-values from \citet{RU} that were corrected by $+0.11$~dex, following the
  recommendation of \citet{Grevesse1999}.

\subsubsection{Codes and model ingredients}\label{Sect:codes}

The present investigation is based on the NLTE methods developed in our earlier studies and described in detail by \citet{mash_fe} for \ion{Fe}{i-ii} and \citet{sitnova_ti} for \ion{Ti}{i-ii}. A comprehensive model atom for iron included, for the first time, not only measured but also predicted energy levels of \ion{Fe}{i}, about 3\,000, in total, and used the most up-to-date radiative data on photoionisation cross sections and transition probabilties. Similar approach was applied to construct model atom for titanium, with more than 3\,600 measured and predicted energy levels of \ion{Ti}{i} and 1\,800 energy levels of \ion{Ti}{ii} and using quantum mechanical photoionization cross sections.
To solve the coupled radiative transfer and statistical equilibrium
(SE) equations, we employed a revised version of the {\sc DETAIL} code
\citep{detail} based on the accelerated lambda iteration (ALI) method
described in \citet{rh91,rh92}. An update of the opacity package in {\sc DETAIL} was presented by \citet{mash_fe}. 

We first calculated the LTE elemental abundances with
the code {\sc WIDTH9}\footnote{\tt http://kurucz.harvard.edu/programs/WIDTH/}
\citep[][modified by Vadim Tsymbal, private
  communication]{2005MSAIS...8...14K}. The NLTE abundances were then derived by
applying the NLTE abundance corrections, $\Delta_{\rm NLTE} =
\eps{NLTE}-\eps{LTE}$. For each line and set of stellar atmospheric parameters,
these corrections were obtained either by interpolation of the pre-computed correction grid of \citet{Mashonkina_dnlte2016} or by direct computation with the
code {\sc LINEC} \citep{Sakhibullin1983}. We verified the consistency of the two codes, {\sc WIDTH9} and {\sc LINEC}, in LTE.

We used the MARCS homogeneous spherical atmosphere models with standard abundances
\citep{Gustafssonetal:2008}, as provided by the MARCS web site\footnote{\tt
  http://marcs.astro.uu.se}. They were interpolated at the necessary $\Teff$,
log~g, and iron abundance [Fe/H], using the FORTRAN-based routine written by Thomas
Masseron and available on the same website.

All our codes treat the radiation transfer in plane-parallel geometry, while
using the model atmospheres calculated in spherically-symmetric geometry. Such
an approach is referred by \citet{2006A&A...452.1039H} to as $s\_p$
(inconsistent), in contrast to the consistent spherical ($s\_s$) approach.
Using lines of \ion{Fe}{i} and \ion{Fe}{ii}, \citet{2006A&A...452.1039H}
evaluated the abundance differences between $s\_p$ and $s\_s$ for solar
metallicity models, varying temperature and surface gravity. All the differences
are negative, independently of whether the minority or majority species is
considered and also independently of the stellar parameters. For example, for
the models $\Teff$ / log~g = 4500 / 1.0 and 5000 / 1.5, the abundance difference
($s\_p$ - $s\_s$) is smaller than 0.02~dex for the lines with an equivalent
width of $W <$ 120~m\AA. Similar calculations, however for very metal-poor
stars, were performed by Ryabchikova~et~al. (2017, in preparation). In line with
\citet{2006A&A...452.1039H}, the resulting ($s\_p$ - $s\_s$) differences are
overall negative and, for each model atmosphere, their magnitude depends only on
the line strength. For example, for the 4780 / 1.06 / $-2.44$ model, ($s\_p$ -
$s\_s$) does not exceed 0.06~dex for the $W <$ 120~m\AA\ lines. Thus, the
sphericity effects on the abundance
differences between \ion{Fe}{i} and \ion{Fe}{ii} are minor. Our spectroscopic determination
of stellar surface gravities is robust.

In a similar homogeneous way, all the codes we used do treat continuum
scattering correctly; i.e., scattering is taken into account not only in the
absorption coefficient, but also in the source function.

\begin{table*} [htbp]
 \caption{\label{Tab:hyd} NLTE abundances of iron in the Sculptor dSph stars computed using accurate \ion{Fe}{i} + \ion{H}{i} rate coefficients and classical Drawinian rates.} 
 \centering
 \begin{tabular}{lccccccccc}
\hline\hline \noalign{\smallskip}
ID & $\Teff$[K] / log~g / [Fe/H] & \multicolumn{2}{c}{LTE} & & \multicolumn{2}{c}{NLTE(Barklem2016)} & & \multicolumn{2}{c}{NLTE(\kH\ = 0.5)} \\
\cline{3-4}
\cline{6-7}
\cline{9-10}\noalign{\smallskip}
   &                         & \ion{Fe}{i} & \ion{Fe}{ii} & \ & \ion{Fe}{i} &  Fe~I~--~Fe~II & \ & \ion{Fe}{i} &  Fe~I~--~Fe~II \\
\noalign{\smallskip} \hline \noalign{\smallskip}
ET0381     & 4570 / 1.17 / $-2.19$ & 5.14 (74) & 5.31 (9) & & 5.27 & $-0.04$ & & 5.23 & $-0.08$ \\
03\_059    & 4530 / 1.08 / $-2.88$ & 4.43 (91) & 4.62 (4) & & 4.66 & ~0.04  & & 4.60 & $-0.02$  \\
07-49      & 4630 / 1.28 / $-2.99$ & 4.46 (22) & 4.59 (4) & & 4.69 & ~0.10 & & 4.64 & ~0.05 \\
074\_02    & 4680 / 1.23 / $-3.06$ & 4.29 (56) & 4.44 (5) & & 4.59 & ~0.15 & & 4.50 & ~0.06 \\
002\_06    & 4390 / 0.68 / $-3.11$ & 4.12 (69) & 4.39 (4) & & 4.35 & $-0.04$ & & 4.33 & $-0.06$ \\
031\_11    & 4670 / 1.13 / $-3.69$ & 3.82 (37) & 3.81 (2) & & 4.20 & ~0.39  & & 4.11 & ~0.30 \\
\noalign{\smallskip}\hline \noalign{\smallskip}
\multicolumn{10}{l}{Numbers in parenthesis indicate the number of lines measured.} 
\end{tabular}
\end{table*}

\subsubsection{Calibration of \kH }\label{Sect:kH}


We now concentrate on the main source of uncertainties in NLTE calculations for
metal-poor stellar atmospheres: the treatment of the inelastic collisions with
the \ion{H}{i} atoms. This study is based on
the \citet{Drawin1968} approximation, as implemented by \citet{Steenbock1984}, with the Drawinian rates scaled by a factor of \kH. It is worth noting that the \ion{H}{i} impact excitation is taken into account also for the forbidden transitions, following \citet{1994PASJ...46...53T} and using a simple relation between hydrogen and electron collisional rates, $C_H = C_e \sqrt{(m_e/m_H)} N_H/N_e$. The same \kH\ value was applied as for the Drawinian rates.
Using slightly different samples of the reference stars, \citet{mash_fe,Bergemann_fe_nlte}, and \citet{2015ApJ...808..148S} estimated \kH\ empirically as 0.1, 1, and 0.5, respectively.

In the present study, we chose to calibrate \kH\ with the seven Sculptor very
metal-poor giants from \paperI\ and \paperII, for which accurate distance-based surface gravities are available.
For each of these stars, the iron abundance has been derived from the \ion{Fe}{i} and
\ion{Fe}{ii} lines under various line-formation assumptions, i.e.,
NLTE conditions with \kH\ = 0.1, 0.5, 1, and under the LTE hypothesis.  We
did not use any strong lines ($W_{obs} >$ 120\,m\AA) in order to
minimize the impact of the uncertainties in both sphericity (see Sect.~\ref{Sect:codes}) and $\Gamma_6$-values on our results.

\begin{figure} 
\hspace{-3mm}  \resizebox{95mm}{!}{\includegraphics{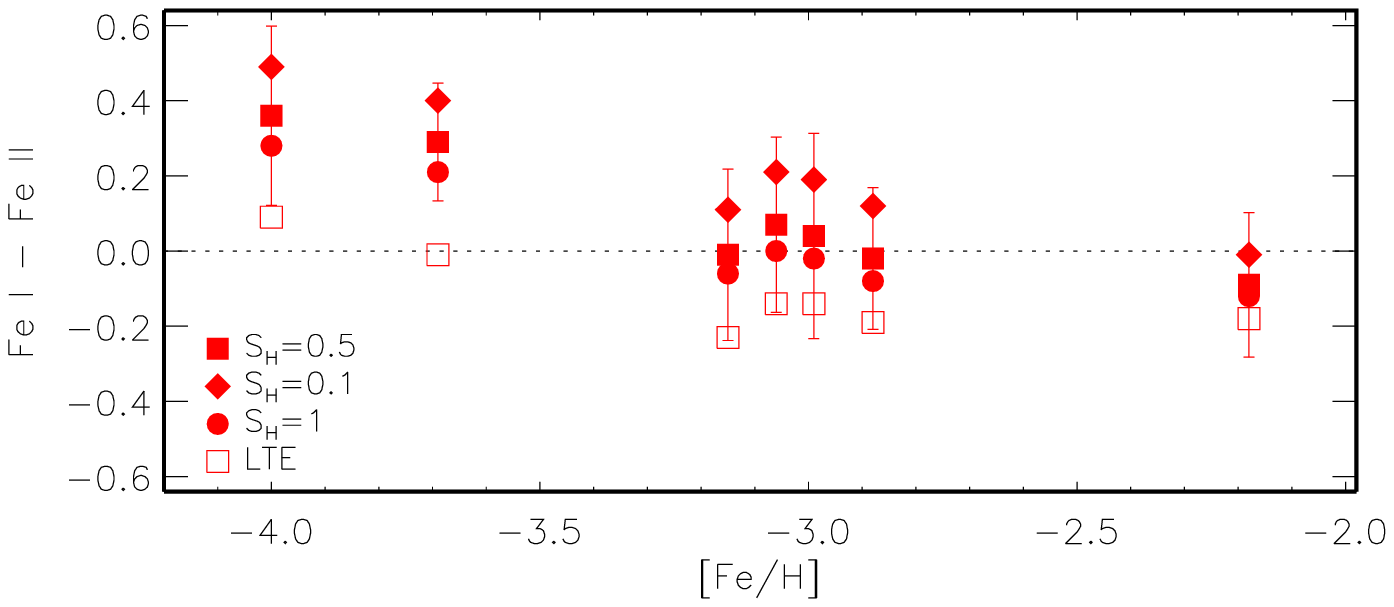}}
  \caption{\label{Fig:fe1_fe2} Abundance differences between the two ionisation stages for iron, Fe~I~--~Fe~II = $\eps{FeI}$ -- $\eps{FeII}$, for the seven Sculptor dSph stars of \paperI\ and \paperII, for the LTE and NLTE line-formation scenarios. The open squares correspond to LTE and the filled rhombi, squares, and circles to NLTE from calculations with \kH\ = 0.1, 0.5, and 1, respectively. The error bars corresponds to $\sigma_{\rm FeI - FeII}$ for NLTE(\kH\ = 0.5).}
\end{figure}

The differences in the mean abundances
derived from lines of \ion{Fe}{i}, $\eps{FeI}$, and \ion{Fe}{ii}, $\eps{FeII}$,
are displayed in Fig.\,\ref{Fig:fe1_fe2}.  
At [Fe/H] $> -3.5$, $\eps{FeI}$ is
systematically lower than $\eps{FeII}$ under the LTE assumption, although the
difference Fe~I~--~Fe~II = $\eps{FeI} - \eps{FeII}$ 
nowhere exceeds $\sigma_{\rm FeI - FeII} = \sqrt{\sigma_{\rm FeI}^2 + \sigma_{\rm FeII}^2}$, which ranges between 0.19~dex and 0.27~dex. Here, the sample standard deviation: $\sigma_{\eps{}} = \sqrt{\Sigma(\overline{x}-x_i)^2 / (N_l-1)}$, determines 
the dispersion in the
single line measurements around the mean for given ionisation stage and $N_l$ is the number of measured lines. For given chemical species, the line-to-line scatter is caused by uncertainties in the continuum normalisation, line-profile fitting (independent of whether in spectral synthesis or equivalent width measurements), and atomic data, and, thus, of random origin. 

\begin{figure} 
  \resizebox{90mm}{!}{\includegraphics{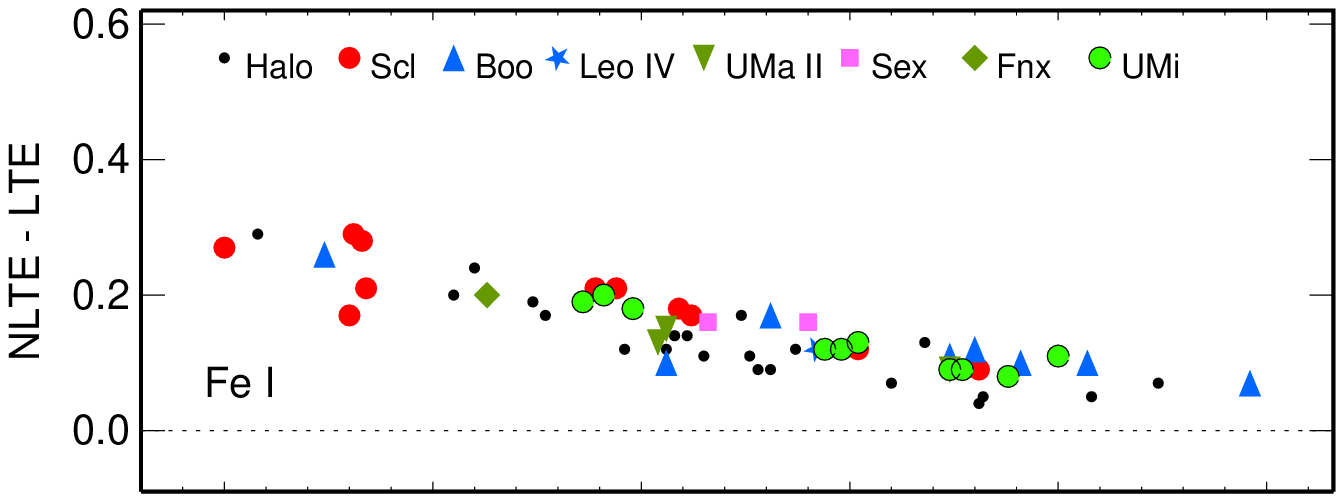}}
  
  \vspace{-13mm}  
  \resizebox{90mm}{!}{\includegraphics{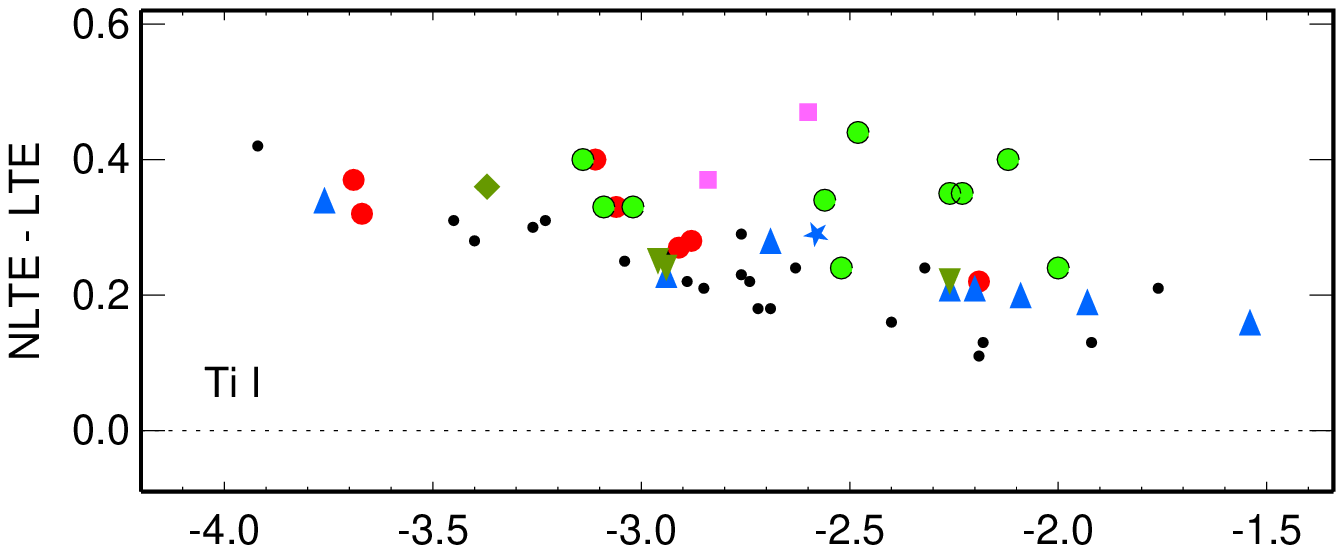}}
  \caption{\label{Fig:dnlte}  Differences between the NLTE and LTE abundance derived from lines of \ion{Fe}{i} (top panel) and \ion{Ti}{i} (bottom panel)   
  in the stars in Sculptor (red circles), Ursa Minor (green circles), Fornax (rhombi), Sextans (squares), Bo{\"o}tes~I (triangles), UMa~II (inverted triangles), and Leo~IV (5 pointed star) dSphs and the MW halo stars (small black circles). 
  }
\end{figure}

Any NLTE treatment results in weaker \ion{Fe}{i} lines as compared to the LTE
approximation. This is due to the overionisation driven by super-thermal radiation
of non-local origin below the ionisation thresholds of the \Eexc\ = 1.4-4.5~eV levels. It therefore
induces positive NLTE abundance corrections, as shown in Fig.\,\ref{Fig:dnlte}. For a given spectral line and model
atmosphere, $\Delta_{\rm NLTE}$ increases with decreasing \kH. At given \kH, the
NLTE effect increases with decreasing metallicity. A thorough discussion of the NLTE abundance corrections for an extended list of the \ion{Fe}{i} lines is given by \citet {mash_fe} and \citet{Mashonkina_dnlte2016}.

At \kH\ = 0.5, $\Delta_{\rm NLTE}$ does not exceed 0.15~dex in the 4570 / 1.17 / $-2.1$ model, while it ranges
between 0.15~dex and 0.45~dex for different lines in the 4670 / 1.13 / $-3.6$ model.
The departures from LTE are small for  \ion{Fe}{ii}, such that $\Delta_{\rm NLTE}$
nowhere exceeds 0.01~dex for \kH\ $\ge$ 0.5 and reaches +0.02~dex
for \kH\ = 0.1 in the most iron-poor models. 

Our test calculations disfavour \kH\ = 0.1 because this leads to higher
abundance from \ion{Fe}{i} than \ion{Fe}{ii} for all stars at the exception of
ET0381, the least metal-poor star of our sample, for which \kH\ = 0.1 leads to
exact identical abundances between the two ionisation stages.

In the [Fe/H] $> -3.5$ regime, \kH\ = 1 leads to somewhat
  negative average difference between \ion{Fe}{i} and \ion{Fe}{ii} ($-0.06\pm0.05$~dex), hence there
  is no reason to increase \kH\ above 0.5, which provides a very satisfactory
  balance between the two ionisation stages. The particular case of our most MP stars,
  which obviously cannot be tackled with \kH, is addressed later in
Section \ref{Sect:emp}.

\begin{figure*}
\resizebox{88mm}{!}{\includegraphics{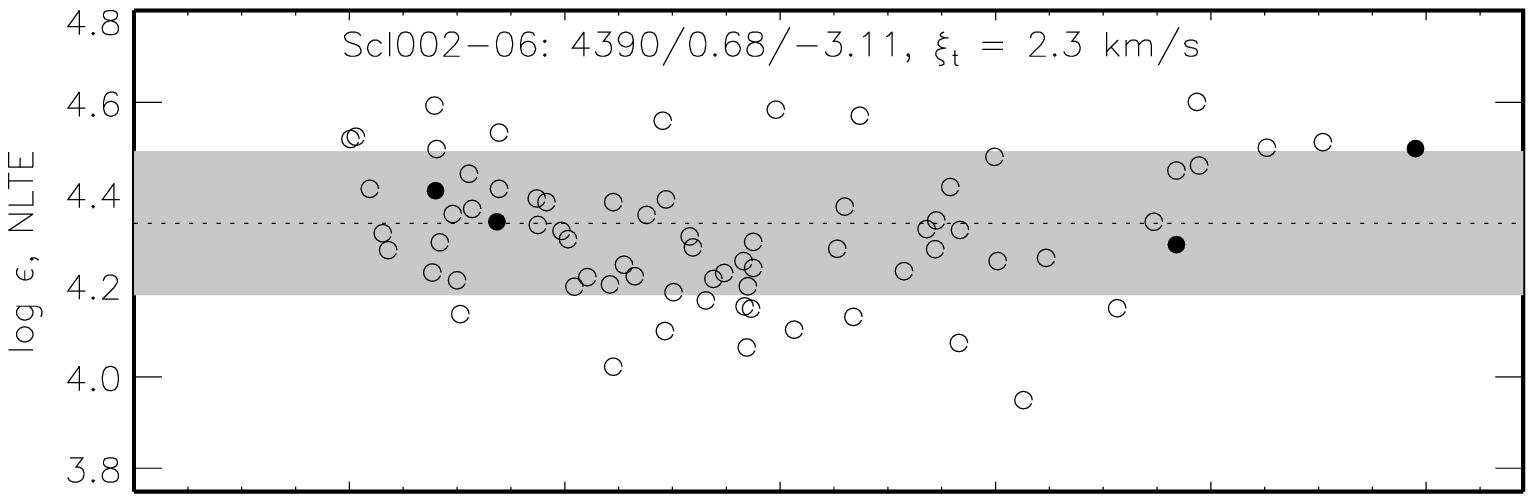}}
\resizebox{88mm}{!}{\includegraphics{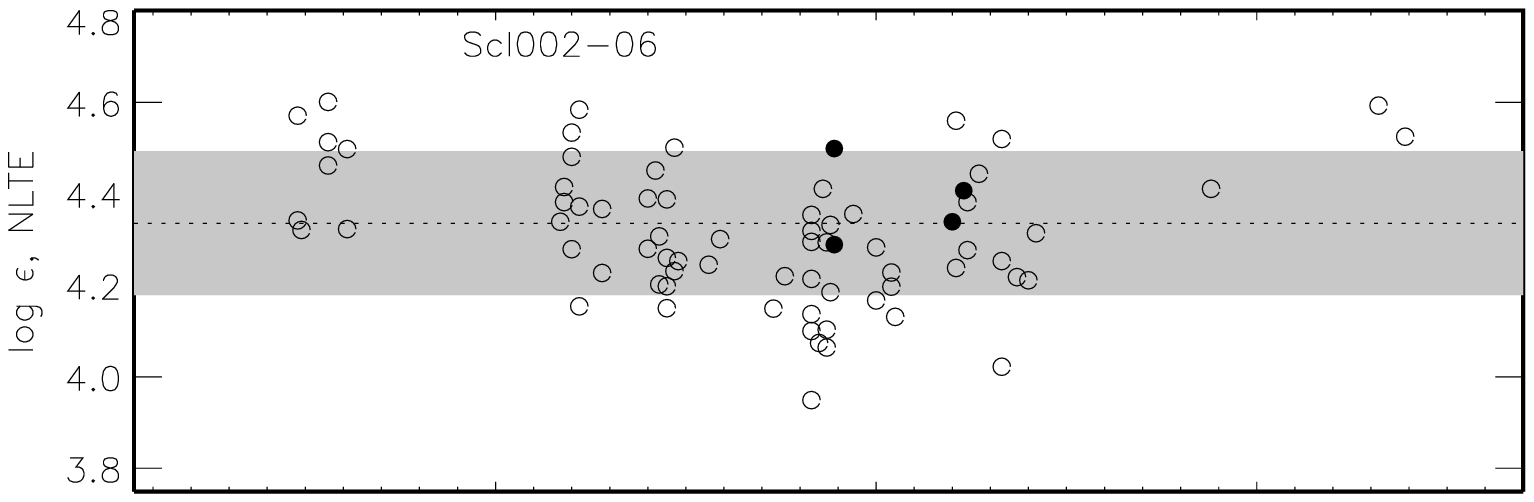}}

  \vspace{-11mm}
\resizebox{88mm}{!}{\includegraphics{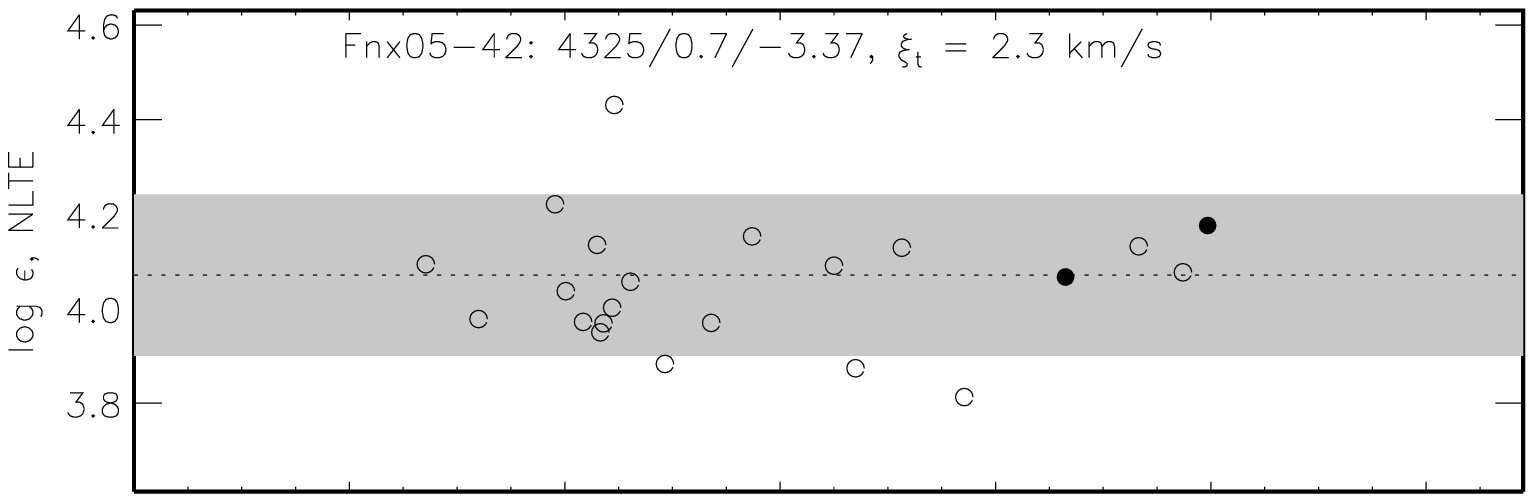}}
\resizebox{88mm}{!}{\includegraphics{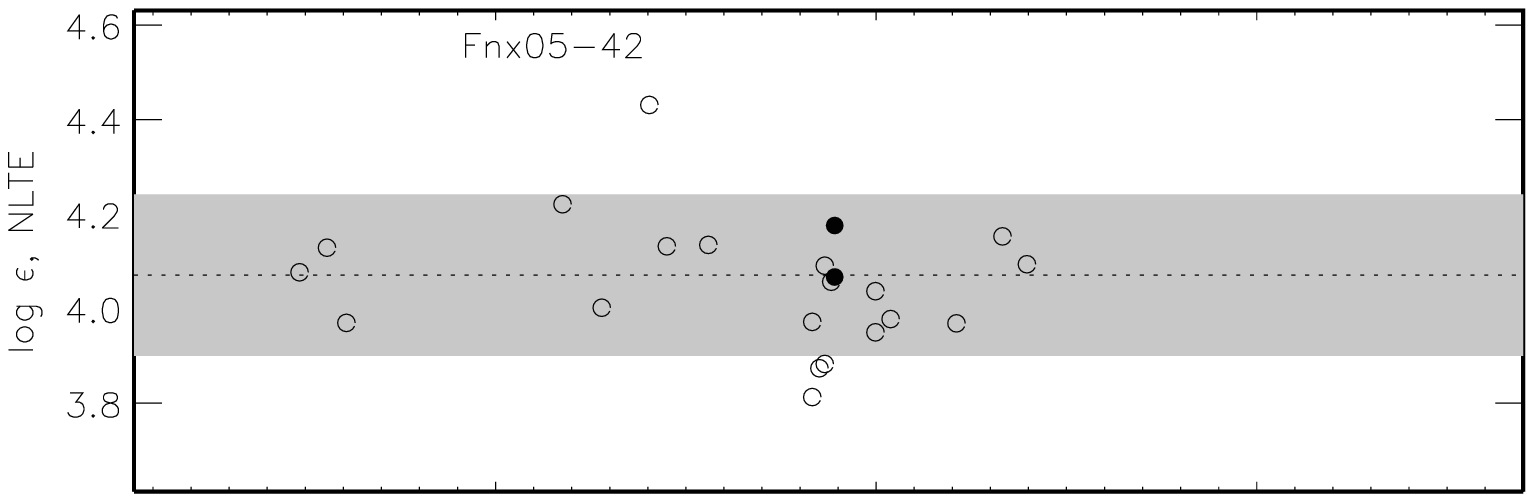}}

  \vspace{-11mm}
\resizebox{88mm}{!}{\includegraphics{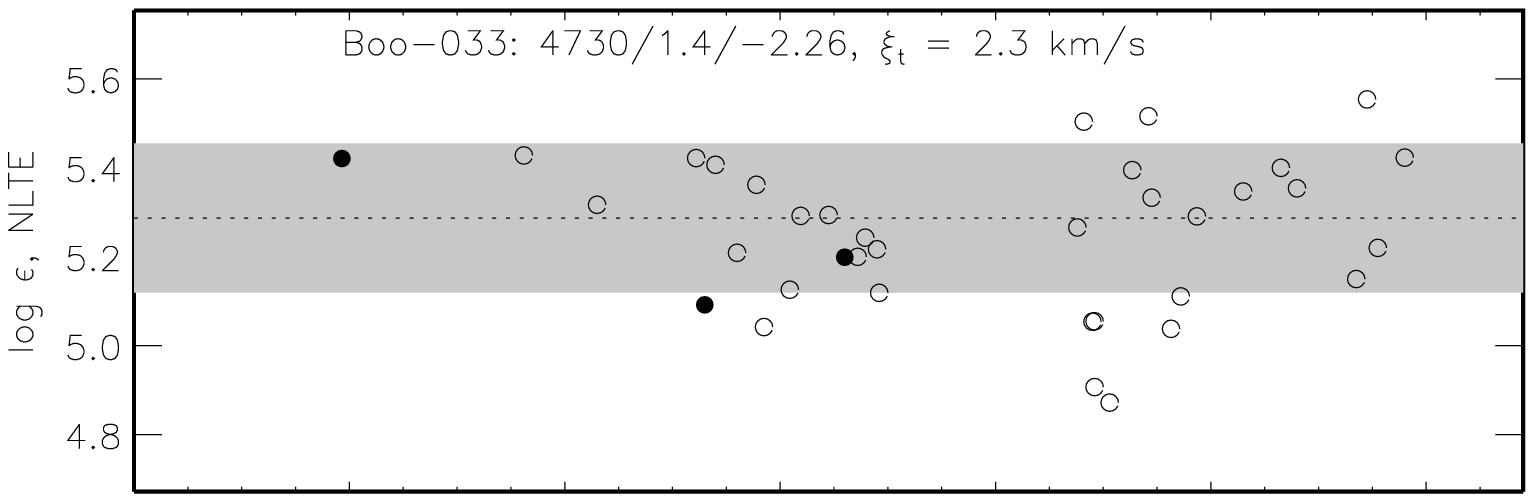}}
\resizebox{88mm}{!}{\includegraphics{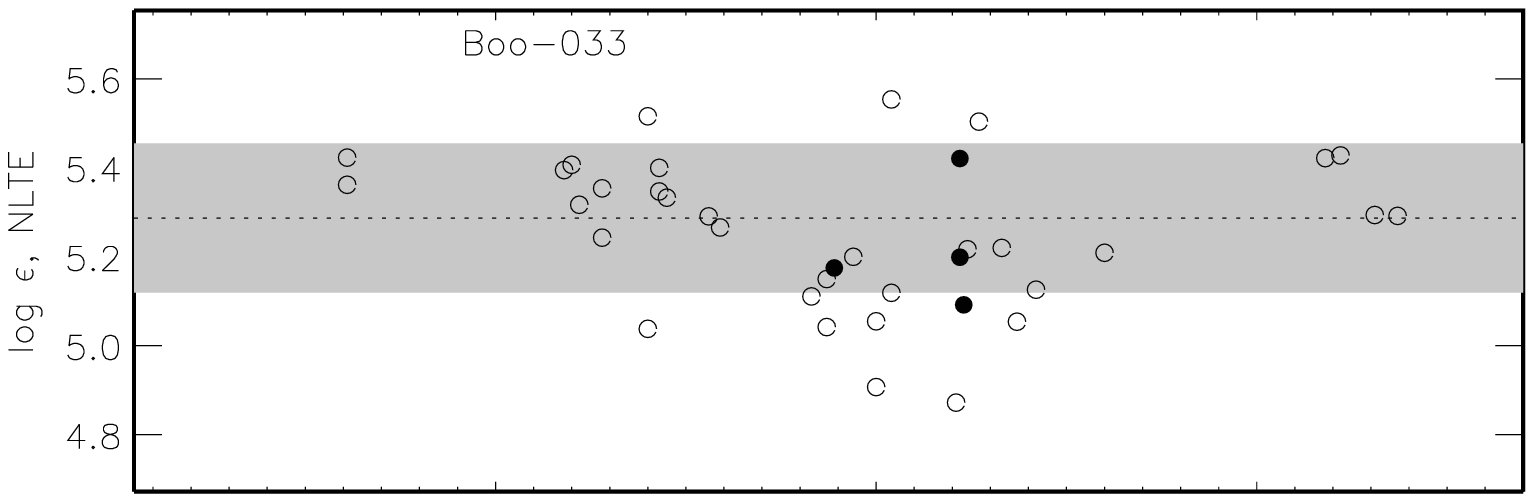}}

  \vspace{-11mm}
\resizebox{88mm}{!}{\includegraphics{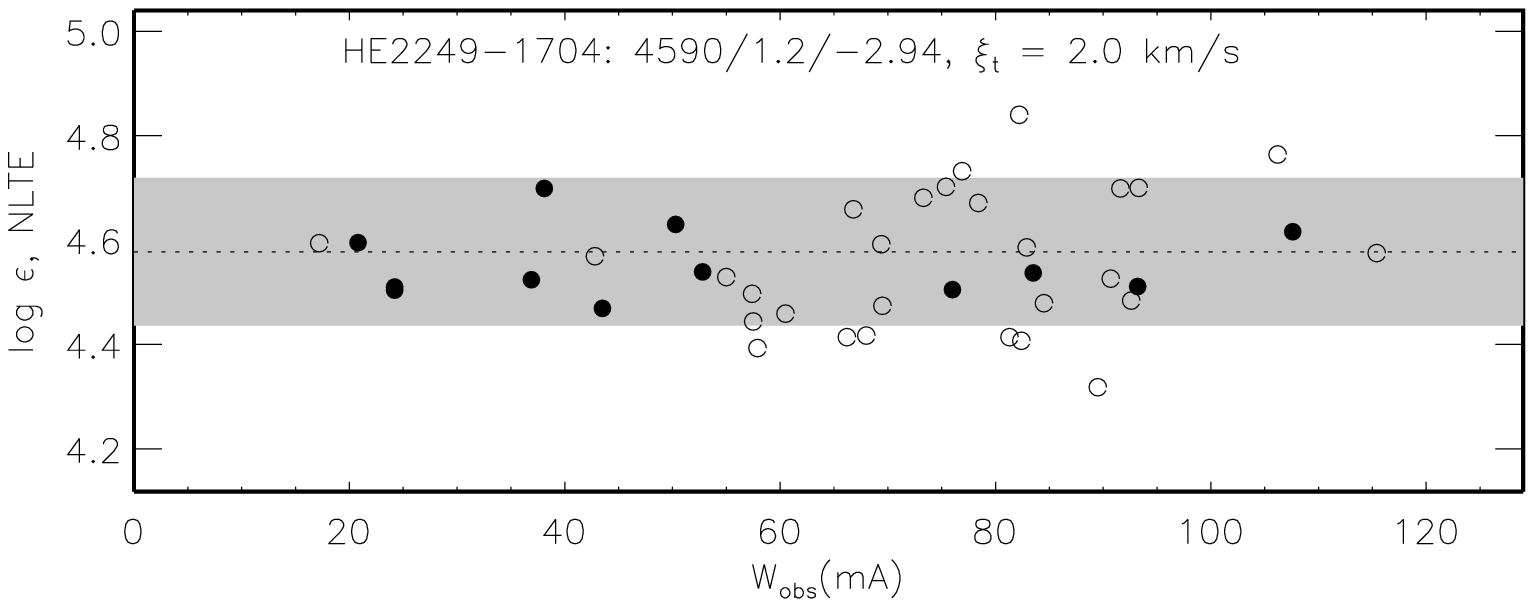}}
\resizebox{88mm}{!}{\includegraphics{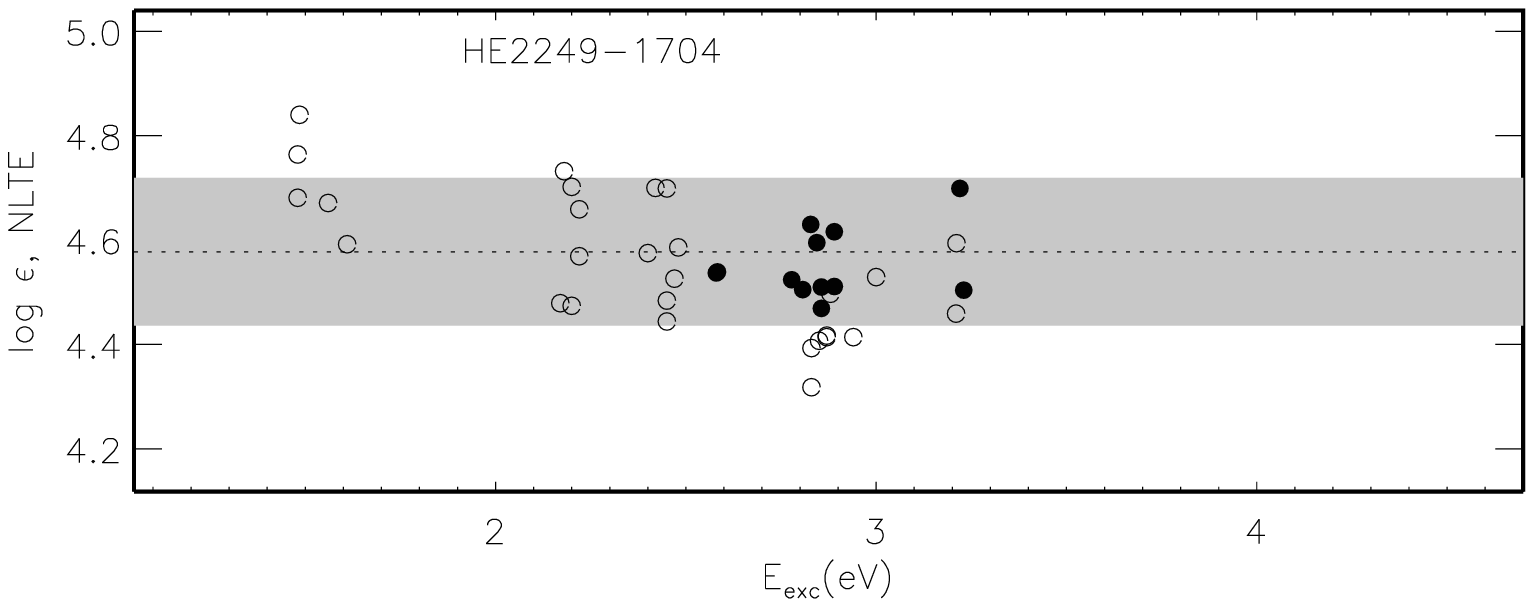}}
\caption[]{NLTE(\kH\ = 0.5) abundances derived from the \ion{Fe}{i} (open circles) and \ion{Fe}{ii} (filled circles) lines in the selected stars as a function of $W_{obs}$ (left column) and \Eexc\ (right column). In each panel, the dotted line shows the mean iron abundance determined from the \ion{Fe}{i} lines and the shaded grey area the scatter around the mean value, as determined by the sample standard deviation.}
\label{Fig:fe_wobs}
\end{figure*}

Not only the \ion{Fe}{i}/\ion{Fe}{ii} ionisation, but also the \ion{Fe}{i}
excitation equilibrium was achieved, when keeping the photometric values of
$\Teff$ and log~g$_d$. Figure\,\ref{Fig:fe_wobs} displays the NLTE (\kH\ = 0.5)
abundances, $\eps{}$, of the individual lines of \ion{Fe}{i} and \ion{Fe}{ii} in
Scl002\_06 and Fnx05-42 as a function of \Eexc\ and $W_{obs}$. These abundances
are put on classical scale with $\eps{H} = 12$. In most cases, NLTE leads to
smaller slopes (in absolute value) than LTE in the relation $\eps{}$(\ion{Fe}{i}) vs E$_{\rm{exc}}$,
for example of $-0.03$~dex/eV instead of $-0.11$~dex/eV for Scl031\_11.

In sharp contrast to the above description, our two most metal-poor stars with
[Fe/H] $< -3.5$ have already \ion{Fe}{i} and \ion{Fe}{ii} consistent
abundances in the LTE approximation. While NLTE leads to Fe~I~--~Fe~II
= 0.21$\pm$0.16~dex for Scl031\_11 and 0.28$\pm$0.24~dex for Scl07-50, even for
\kH\ = 1. At face value, the \ion{Fe}{ii} abundance relies on only two lines, at
4923\,\AA\ and 5018\,\AA, with rather uncertain $gf$-values. Nevertheless, we note
that decreasing $\Teff$ by 170~K and 200~K for Scl031\_11 and Scl07-50,
respectively, leads to consistent NLTE iron abundances from the two
ionisation stages, when adopting \kH\ = 0.5.

\citet{2016PhRvA..93d2705B} has treated a theoretical method for the estimation of cross sections and rates for excitation and charge-transfer processes in low-energy hydrogen-atom collisions with neutral atoms, based on an asymptotic two-electron model of ionic-covalent interactions in the neutral atom-hydrogen-atom system and the multichannel Landau-Zener model. The rate coefficients computed for \ion{Fe}{i}+\ion{H}{i} collisions were applied by \citet{2016MNRAS.tmp.1203A,2017A&A...597A...6N}, and \citet{2017arXiv170304027L} to the NLTE analyses of lines of iron in the reference metal-poor stars. Paul Barklem has kindly provided us with the \ion{Fe}{i}+\ion{H}{i} rate coefficients, and we applied these data to determine the iron NLTE abundances of the six stars in the Sculptor dSph. Following \citet{2016MNRAS.tmp.1203A}, inelastic collisions of \ion{Fe}{ii} with \ion{H}{i} were treated using the scaled Drawinian rates. We adopt \kH\ = 0.5. The obtained results are presented in Table\,\ref{Tab:hyd}.
For \ion{Fe}{i}, the NLTE~--~LTE abundance difference ranges between 0.13~dex and 0.38~dex, depending of the star's metallicity. In the four stars, NLTE leads to acceptable abundance difference of no more than 0.10~dex between \ion{Fe}{i} and \ion{Fe}{ii}. However, implementing the most up-to-date \ion{Fe}{i} + \ion{H}{i} collision data in our NLTE model does not help to  achieve the \ion{Fe}{i}/\ion{Fe}{ii} ionisation equilibrium for the [Fe/H] $\simeq -3.7$ star.

\subsubsection{Determination of log~g from analysis of \ion{Fe}{i}/\ion{Fe}{ii}}\label{sect:sp}

Having realised that the \citet{Drawin1968} approximation does not contain the relevant physics \citep[see, for example, a critical analysis of][]{Barklem2011_hyd} and, for different transitions in \ion{Fe}{i}, the Drawinian rate has a different relation to a true \ion{Fe}{i}+\ion{H}{i} collision rate, 
we consider the NLTE calculations with the scaled Drawinian rates in the 1D model atmospheres as an 1D-NLTE(\kH\ = 0.5) model that fits observations of the \ion{Fe}{i} and \ion{Fe}{ii} lines in our reference stars, namely, the Sculptor dSph stars with [Fe/H] $> -3.7$. This model was tested further with our stellar samples in the Ursa Minor, Fornax, Sextans, Bo\"otes~I, Leo~IV, and UMa~II dSphs. The \ion{Fe}{i}/\ion{Fe}{ii} ionisation
equilibrium was checked in each star, while keeping its atmospheric parameters, $\Teff$, log~g$_d$ or log~g$_{ph}$, fixed.
From there we determined the final iron
abundances and the microturbulence velocites, $\xi_t$. They are presented in Table\,\ref{Tab:parameters}. For example, we show in Fig.\,\ref{Fig:fe_wobs} the NLTE abundances from lines of \ion{Fe}{i} and \ion{Fe}{ii} in Boo-33 as functions of $W_{obs}$ and \Eexc, which support the derived $\xi_t$ = 2.3~\kms\ and $\Teff$ = 4730~K.  
Noteworthily, we did not find any significant change in the slopes of the
$\eps{}$(\ion{Fe}{i}) -- log($W_{\lambda}/\lambda$) plots between the LTE and NLTE
calculations.

Table\,\ref{Tab:fe_ti} lists the mean NLTE abundances from each ionisation stage, \ion{Fe}{i} and \ion{Fe}{ii}, together with their $\sigma_{\eps{}}$ and number of lines measured. 
Systematic errors of $\eps{FeI}$ and $\eps{FeII}$ for a given star are due to the uncertainty in adopted atmospheric parameters. Our calculations show that a change of +100~K in $\Teff$ produces 0.10-0.12~dex higher abundances from lines of \ion{Fe}{i} and has minor effect ($<$ 0.02~dex) on the abundances from lines of \ion{Fe}{ii}. In contrast, a change of 0.1~dex in log~g has minor effect ($< 0.01$~dex) on \ion{Fe}{i} and shifts $\eps{FeII}$ by +0.04~dex. A change of +0.2\,\kms\ in $\xi_t$ produces 0.02~dex to 0.05~dex lower iron abundances, depending on the sample of the iron lines measured in given star.

\begin{figure} 
\hspace{-3mm}  \resizebox{95mm}{!}{\includegraphics{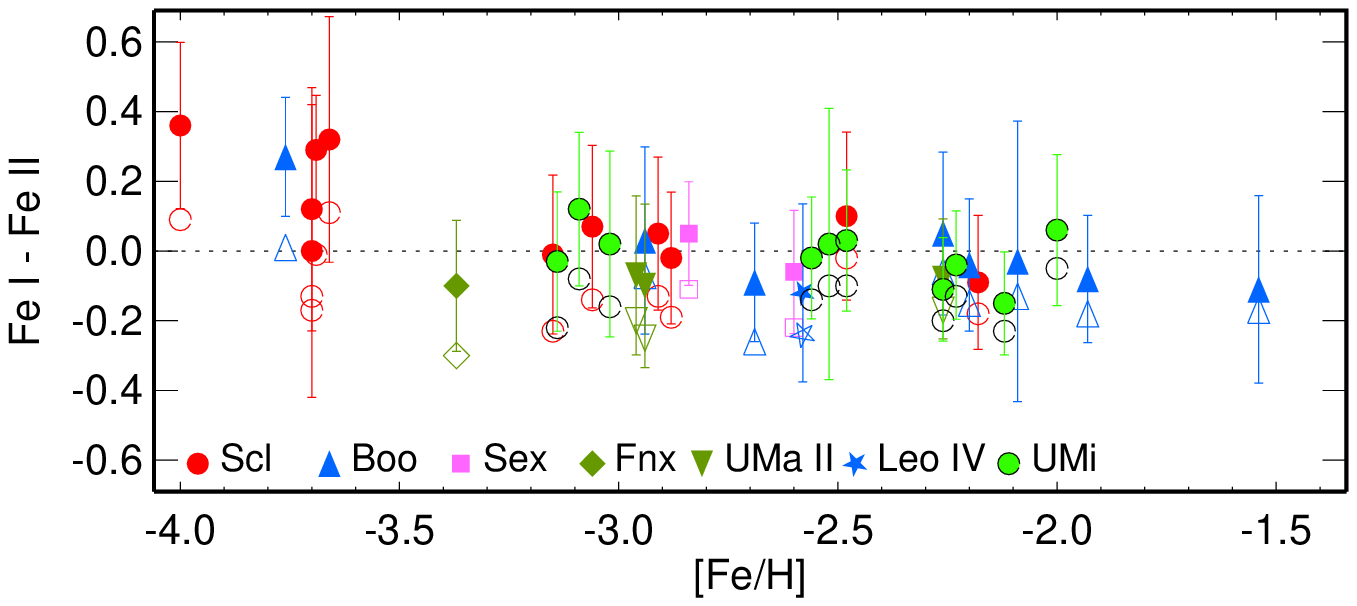}}

\vspace{-5mm}
\hspace{-3mm}  \resizebox{95mm}{!}{\includegraphics{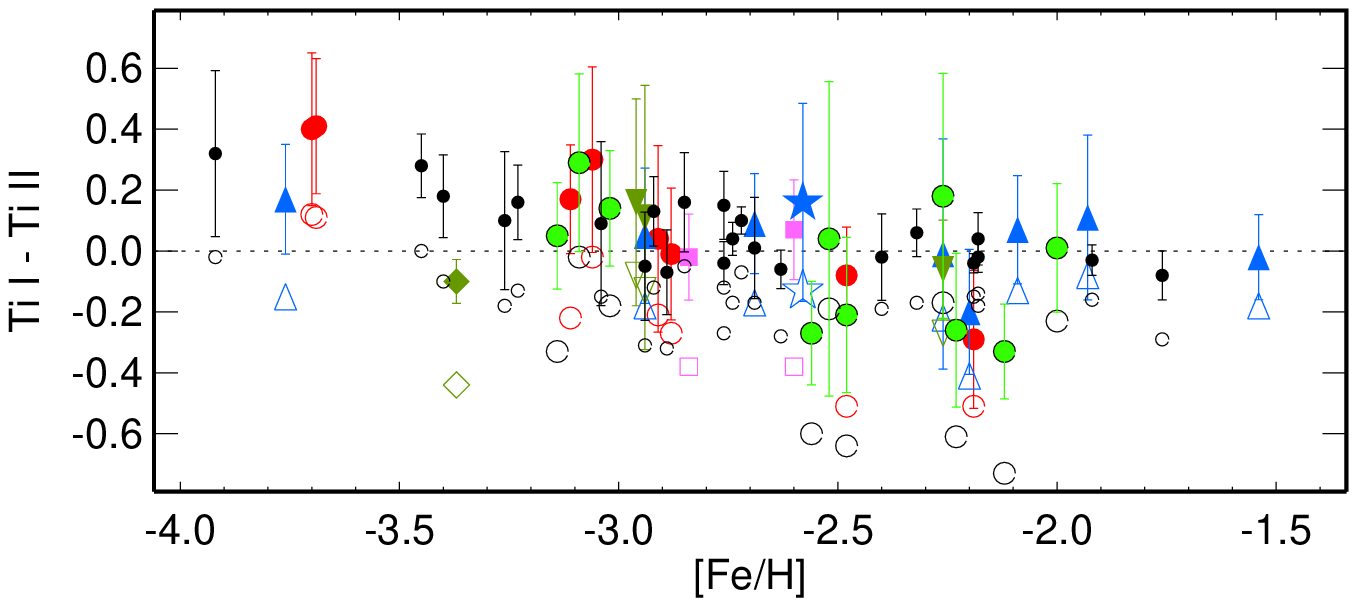}}
  \caption{\label{Fig:fe1_fe2_all} Abundance differences between the two ionisation stages of iron, Fe~I -- Fe~II (top panel), and titanium, Ti~I -- Ti~II (bottom panel), in the investigated stars in Sculptor, Ursa Minor, Fornax, Sextans, Bo{\"o}tes~I, UMa~II, and Leo~IV dSphs. For the MW halo stars only Ti~I -- Ti~II is shown. Symbols as in Fig.~\ref{Fig:dnlte}. The open and filled symbols correspond to the LTE and NLTE line-formation scenario, respectively. Here, \kH\ = 0.5 for \ion{Fe}{i-ii} and \kH\ = 1 for \ion{Ti}{i-ii}. }
\end{figure}

The LTE and NLTE(\kH\ = 0.5)
abundance differences between \ion{Fe}{i} and \ion{Fe}{ii} are shown in the
upper panel of Fig.\,\ref{Fig:fe1_fe2_all}.
In addition to Scl07-50 and Scl031\_11, another two stars, Scl~6\_6\_402 ([Fe/H] = $ -3.66$) and Boo-1137 ([Fe/H] = $-3.76$), have differences of more than 0.2~dex in the NLTE abundances between the two ionisation stages.
Other than these, the average $\eps{FeI}$ - $\eps{FeII}$ difference amounts in NLTE to $-0.02\pm0.07$.
It is worth noting, we have the two [Fe/H] $\simeq -3.7$ stars, Scl11\_1\_4296 and S1020549, for which the \ion{Fe}{i}/\ion{Fe}{ii} ionisation equilibrium is fulfilled in NLTE.
Hence, our 1D-NLTE(\kH\ = 0.5) model can reliably be used to
determine the spectroscopic gravity of stars with [Fe/H] $\succsim -3.7$. Consequently, we adopted
as final value the spectroscopic log~g$_{sp}$ = 1.8 instead of log~g$_d$ = 1.63
for Boo-980 (4760 / 1.8 / $-3.01$).

Based on the NLTE analysis of the \ion{Fe}{i} and \ion{Fe}{ii} lines, we checked $\Teff$ / log~g determined by \citet{Cohen2013} for their MW stellar subsample. The surface gravities
were revised by +0.1~dex to +0.2~dex, well within 1$\sigma$ uncertainties. 
The obtained microturbulence velocities are similar to those of \citet{Cohen2013} except for the four stars, for which our values are 0.2~\kms\ to 0.5~\kms\ lower. One of them is HE2249-1704 (4590 / 1.2 / $-2.94$), and Fig.\,\ref{Fig:fe_wobs} supports the derived $\xi_t$ = 2~\kms.
It is worth noting, we did not include the \ion{Fe}{ii} 3255\,\AA, 3277\,\AA, and 3281\,\AA\ lines in
the analysis of BS16550-087 because they give 0.81~dex, 0.27~dex, and 0.44~dex
higher abundances than the mean of the other twelve \ion{Fe}{ii} lines. 

For the rest of the MW stellar sample, their log~g, iron abundance, and microturbulence velocity were determined in this study from the requirements that (i) the NLTE abundances from \ion{Fe}{i} and \ion{Fe}{ii} must be equal and (ii) lines of \ion{Fe}{i} with different equivalent widths must yield equal NLTE abundances.

\begin{table*} [htbp]
 \caption{\label{Tab:error} Error budget for log~g$_{sp}$ of HD~8724 and HE1356-0622.} 
 \centering
 \begin{tabular}{lrrrcrrr}
\hline\hline \noalign{\smallskip}
 Source of & \multicolumn{3}{c}{HD~8724} & \ & \multicolumn{3}{c}{HE1356-0622} \\
 uncertainty & \multicolumn{3}{c}{4560/1.29/$-1.76$, $\xi_t$ =  1.5\,\kms} & & \multicolumn{3}{c}{4945/2.00/$-3.45$, $\xi_t$ = 2.0\,\kms} \\
\cline{2-4}
\cline{6-8}
 \noalign{\smallskip}
 & $\eps{FeI}$ & $\eps{FeII}$ & log~g & & $\eps{FeI}$ & $\eps{FeII}$ & log~g \\
\noalign{\smallskip} \hline \noalign{\smallskip}
Line-to-line scatter & $\pm$0.09 & $\pm$0.03 & $\pm$0.19 & & $\pm$0.12 & $\pm$0.08 & $\pm$0.25 \\
$\Delta\Teff$ = 50~K & 0.07      & $<$ 0.01  & 0.15      & & 0.05      & $<$ 0.01  & 0.10 \\
NLTE model           & $-$0.05   & 0.0       & $-$0.1    & & $-$0.11   & 0.0       & $-$0.23 \\
\noalign{\smallskip}\hline \noalign{\smallskip}
\end{tabular}
\end{table*}

Our calculations show that a change of 0.1~dex in $\eps{FeI}$~--~$\eps{FeII}$ leads to a shift of 0.23~dex and 0.19~dex in log~g for the model atmospheres 4945/2.00/$-3.45$ and 4560/1.29/$-1.76$, respectively. Table\,\ref{Tab:error} shows estimates of the random and systematic errors, $\sigma_{\rm log g(Sp)}$ and $\Delta_{\rm log g(Sp)}$, of the derived spectroscopic surface gravity for the two stars, HE1356-0622 and HD~8724, which represent the most and the least metal-poor samples. The random error is contributed from the line-to-line scatter for \ion{Fe}{i} and \ion{Fe}{ii} that is represented by $\sigma_{\rm FeI}$ and $\sigma_{\rm FeII}$ and the uncertainty in $\Teff$. The quadratic addition of the individual uncertainties results in $\sigma_{\rm log g(Sp)}$ = 0.24 and 0.32 for HD~8724 and HE1356-0622, respectively. 

Uncertainty in the NLTE model was assumed to be mostly produced by applying the scaled (\kH\ = 0.5) Drawinian rates instead of quantum-mechanical rate coefficients of \citet{2016PhRvA..93d2705B}. In all cases, this leads to less positive NLTE abundance corrections for \ion{Fe}{i} and, thus, to systematically underestimated surface gravity. For our sample of the MW giants, $\Delta_{\rm log g(Sp)}$ ranges between $-0.1$~dex and $-0.23$~dex.

The final atmospheric parameters are presented in Table\,\ref{Tab:parameters}.
The iron abundance is defined from lines of \ion{Fe}{ii}. For the
computation of the abundances relative to the solar scale we employed
$\eps{Fe,\odot}$ = 7.50 \citep{1998SSRv...85..161G}.

\section{Checking atmospheric parameters with independent methods}\label{Sect:test}

\subsection{\ion{Ti}{i}/\ion{Ti}{ii} ionisation equilibrium}\label{Sect:ti}

The fact that for most of our stars, titanium is accessible in the two
ionisation stages, \ion{Ti}{i} and \ion{Ti}{ii}, opens another opportunity to
check the stellar surface gravities.

We used accurate and homogeneous $gf$-values of the \ion{Ti}{i} and \ion{Ti}{ii}
lines from laboratory measurements of \citet{Lawler2013_ti1} and
\citet{2013_gf_ti2}. The LTE and NLTE abundance differences between \ion{Ti}{i}
and \ion{Ti}{ii} are displayed in Fig.\,\ref{Fig:fe1_fe2_all}.

In LTE, lines of \ion{Ti}{i} systematically give lower abundances, by up to 0.51~dex,
than the \ion{Ti}{ii} lines. The only exception is S1020549, with two weak
($W_{obs} \simeq$ 25\,m\AA) lines of \ion{Ti}{i} measured in its R $\simeq$
33\,000 spectrum.

Similarly to \ion{Fe}{i}, the main NLTE mechanism for \ion{Ti}{i} is the UV
overionisation, resulting in weakened lines and positive NLTE abundance
corrections (Fig.\,\ref{Fig:dnlte}). Since there is no accurate data on inelastic collisions of the
titanium atoms with \ion{H}{i}, we rely on the Drawinian rates (i.e. \kH\ = 1), as
recommended by \citet{sitnova_ti}.  
For a given stellar atmosphere model, the NLTE corrections of the individual
\ion{Ti}{i} lines are of very similar orders of magnitude. For example,
$\Delta_{\rm NLTE}$ ranges between 0.17~dex and 0.21~dex in the 5180 / 2.70 / $-2.60$
model. The departures from LTE grow towards lower metallicity, as shown in
Fig.\,\ref{Fig:dnlte}.

The NLTE corrections for the \ion{Ti}{ii} lines are much smaller than those obtained
for \ion{Ti}{i} and mostly positive. They are close to 0 at [Fe/H] $> -2.5$, but
increase with decreasing metallicity and are close to $\Delta_{\rm NLTE}$ =
0.1~dex at [Fe/H] = $-4$.

At [Fe/H] $> -3.5$, NLTE leads to consistent abundances between \ion{Ti}{i} and \ion{Ti}{ii} in most
of the stars, at the exception of ET0381, for which $\eps{TiI} - \eps{TiII}$ =
$-0.32$~dex. This means that larger NLTE correction is needed for \ion{Ti}{i},
i.e. \kH\ $< 0.1$, to achieve the \ion{Ti}{i}/\ion{Ti}{ii} ionisation equilibrium.

In contrast, the LTE assumption is working well for the [Fe/H] $< -3.5$ stars,
while NLTE  worsen the results. This is similar to what we had found for
\ion{Fe}{i}/\ion{Fe}{ii}. Again a decrease of $\Teff$ by 170~K  for Scl031\_11 removes, partly, a
discrepancy between \ion{Ti}{i} and \ion{Ti}{ii}, however, the remaining
difference of 0.20~dex is still large.

\subsubsection{The specific case of the [Fe/H] $< -3.5$ stars}\label{Sect:emp}

The above analysis reflects that 
the four EMP stars in the dSphs do not have a
  satisfactory ionisation balance between \ion{Fe}{i} and
\ion{Fe}{ii} and between \ion{Ti}{i} and \ion{Ti}{ii}, under the same conditions as the other
  stars. 
Are we facing here problems with the NLTE
line-formation modelling and lack of thermalising processes?  Are we missing a
proper 3D treatment of the stellar atmospheres? Should we simply revise downward
the effective temperature of these EMP stars?

As shown in Sect.~\ref{Sect:kH}, implementing the most up-to-date \ion{Fe}{i} + \ion{H}{i} collision data in our NLTE model does not help to remove the abundance discrepancy between \ion{Fe}{i} and \ion{Fe}{ii}.
Accurate calculations of the \ion{Ti}{i} + \ion{H}{i} collisions would be highly desirable.

Next is the 3D effects.
For \ion{Fe}{i}, we ignored the lines with \Eexc\ $<$ 1.2~eV. This
may partly explain why we obtained smaller difference between \ion{Fe}{i} and
\ion{Fe}{ii} than between \ion{Ti}{i} and \ion{Ti}{ii}. Still, the difference
between $\eps{FeI}$ and $\eps{FeII}$ 
ranges in NLTE (assuming \kH\ = 0.5) between 0.30~dex and 0.36~dex for our four [Fe/H] $\precsim -3.7$ stars. This is
not negligible and can unlikely be removed by the 3D-NLTE calculations. Indeed,
recent papers of \citet{2017A&A...597A...6N} and \citet{2016MNRAS.tmp.1203A}
show that the 3D effects for
\ion{Fe}{i} are of different sign in NLTE than in LTE: 3D-1D = +0.20~dex in NLTE
and $-0.34$~dex in LTE in the 5150 / 2.2 / $-5$ model \citep{2017A&A...597A...6N}
and 3D-1D = +0.11~dex in NLTE and $-0.11$~dex in LTE in the 6430 / 4.2 / $-3$ model
\citep{2016MNRAS.tmp.1203A}. In the latter model, abundance from lines of
\ion{Fe}{ii} is higher in 3D-NLTE than 1D-NLTE. As a result, $\eps{FeI}$ - $\eps{FeII}$ is
0.05~dex smaller in 3D-NLTE than 1D-NLTE. No data is provided on \ion{Fe}{ii} in
the 5150 / 2.2 / $-5$ model. It would be important to perform the 3D-NLTE
calculations for \ion{Fe}{i-ii} in the [Fe/H] = $-4$ model.

For the titanium lines in the red giant atmospheres, the 3D effects were
predicted under the LTE assumption by \citet{2013A&A...559A.102D}. For \ion{Ti}{i},
the (3D-1D) abundance corrections are negative, with a magnitude depending
strongly on \Eexc. For example, in the 5000 / 2.5 / $-3$ model, (3D-1D) =
$-0.45$~dex and $-0.10$~dex for the \Eexc\ = 0 and 2~eV lines at $\lambda$ =
4000~\AA. The 3D effects are predicted to be minor for \ion{Ti}{ii}, with
either positive or negative (3D-1D) correction of less than 0.07~dex in absolute
value. Since our abundance analysis of the EMP stars is based on the
low-excitation (\Eexc\ $\le$ 0.85~eV) lines of \ion{Ti}{i}, a 3D treatment (if
NLTE follows LTE, see below) might help to reconcile $\eps{TiI}$ and $\eps{TiII}$.

At this stage, we conclude that, most likely, the problem we see in NLTE with the
\ion{Fe}{i}/\ion{Fe}{ii} and \ion{Ti}{i}/\ion{Ti}{ii} ionisation equilibrium in
our most MP stars is related to the $\Teff$ determination, given the fact that
the colour calibrations we used are in fact valid in the metallicity range $-3.5
\le$ [Fe/H] $\le$ 0.4 \citep{2005ApJ...626..465R}.

\begin{table} 
 \caption{\label{Tab:gaia} Comparison of the derived spectroscopic surface gravities with that based on the Gaia parallaxes.}
 \centering
 \begin{tabular}{lcccc}
\hline\hline \noalign{\smallskip}
 Star  & $\pi_{Gaia}$[mas] & log~g$_{Gaia}$ & log~g$_{sp}$ \\
\noalign{\smallskip} \hline \noalign{\smallskip}
HD~2796   & 1.64$\pm$0.26 & 1.84$\pm$0.12 & 1.55 \\
HD~4306   & 1.78$\pm$0.41 & 2.16$\pm$0.18 & 2.18 \\
HD~8724   & 2.84$\pm$0.27 & 2.05$\pm$0.08 & 1.29 \\
HD~218857 & 3.03$\pm$0.25 & 2.64$\pm$0.06 & 2.53 \\
BD $-11^\circ$0145 & 0.56$\pm$0.24 & 1.81$\pm$0.31 & 1.73 \\
\noalign{\smallskip}\hline \noalign{\smallskip}
\end{tabular}
\end{table}

\subsection{Spectroscopic versus Gaia DR1 gravities}

As a sanity check, we computed the distance-based log~g for the five stars with
available Gaia parallax measurements \citep[][Gaia Data Release~1]{2016A&A...595A...2G} and available in the VizieR
Online Data Catalog. As can be seen in Table\,\ref{Tab:gaia}, for three stars,
our log~g$_{sp}$ are consistent with log~g$_{Gaia}$ within the error bars. This
holds despite the fact that two of these stars, HD~4306 and BD $-11^\circ$0145,
are identified as binaries.

There is one exception to this general agreement, HD~8724. It is hard to
understand a source of an extremely large difference of 0.76~dex(!) between
log~g$_{Gaia}$ and log~g$_{sp}$, in view of their small statistical and systematic errors, i.e., $\sigma_{\rm log g(Gaia)}$ = 0.08~dex, $\sigma_{\rm log g(sp)}$ = 0.24~dex and $\Delta_{\rm log g(sp)}$ = $-0.1$~dex. The effective temperature of HD~8724 should be
increased by $\sim$400~K in order to reconcile the \ion{Fe}{i} and \ion{Fe}{ii}
abundances with log~g$_{Gaia}$ = 2.05. This seems very unlikely. All estimates,
based on the infrared flux method (IRFM) are close to $\Teff$ = 4560~K derived in
this study: $\Teff$ = 4535~K \citep{1999A&AS..139..335A},
4540~K \citep{2005ApJ...626..446R}, and 4630~K \citep{2009A&A...497..497G}.

\subsection{Checking atmospheric parameters with evolutionary tracks}

\begin{figure}
 \resizebox{90mm}{!}{\includegraphics{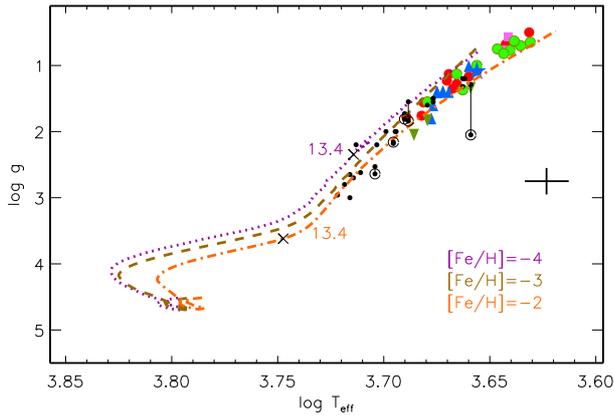}}
 \caption[]{Investigated stars compared with the evolutionary tracks of $M$ = 0.8\,$M_\odot$ and [Fe/H] = $-2$ (dash-dotted curve), $-3$ (dashed curve), and $-4$ (dotted curve). The crosses on the [Fe/H] = $-2$ and [Fe/H] = $-4$ evolutionary tracks mark stellar age of 13.4~Gyr. Symbols as in Fig.~\ref{Fig:dnlte}. For the five MW stars with the Gaia parallax available the vertical lines connect the star's positions corresponding to log~g$_{sp}$ (small black circles) and log~g$_{Gaia}$ (small black circles inside larger size circles). The cross in the right part indicates log~g and $\Teff$ error bars of 0.2~dex and 100~K, respectively. }
\label{Fig:isochrones}
\end{figure}

We now check the effective temperatures and surface gravities that we derived by
looking at the positions of the stars in the log~g - $\Teff$ diagram. For this
we consider the theoretical $\alpha$-enhanced ([$\alpha$ / Fe] = 0.6) evolutionary
tracks of \citet{Yi2004}. Consistently with our calculation, we assumed the
stellar masses to be 0.8~M$_\odot$.  Figure\,\ref{Fig:isochrones} shows that all
 stars correctly sit on the giant branch between the evolutionary tracks of
[Fe/H] = $-2$ and $-4$, in line with their metallicities.

Very metal-poor stars in the MW halo and dSphs do not exactly span the same
log~g, $\Teff$ range. This comes as a consequence of the observational
constraints. The dSphs are obviously more distant, and their stars are fainter,
hence one tends to target the tip of the RGB. Since NLTE corrections depend on
the stellar atmosphere parameters, in a way which itself depends on the
species, any valuable comparison between different galaxies should be done via
NLTE homogeneous analysis.

For each of five MW stars with available Gaia parallaxes,
Fig.\,\ref{Fig:isochrones} indicates the two positions corresponding to
log~g$_{sp}$ and log~g$_{Gaia}$. With log~g$_{Gaia}$ = 2.05, HD~8724 lies far
from the [Fe/H] = $-2$ evolutionary track. Obviously, the parallax of HD~8724
needs to be revised.

\subsection{Approximate formula for microturbulence value}

The relation between the microturbulence velocities and the basic atmospheric
parameters $\Teff$, log~g, and [Fe/H] is not particularly well established. We
take the opportunity of this study to derive an empirical formula, which we
hope can be useful:

\begin{equation}
\xi_t = 0.14 - 0.08 \times {\rm [Fe/H]} + 4.90 \times (\Teff / 10^4) - 0.47 \times {\rm log~g}. \label{formula1}
\end{equation}

\begin{figure} 
\hspace{-3mm}  \resizebox{95mm}{!}{\includegraphics{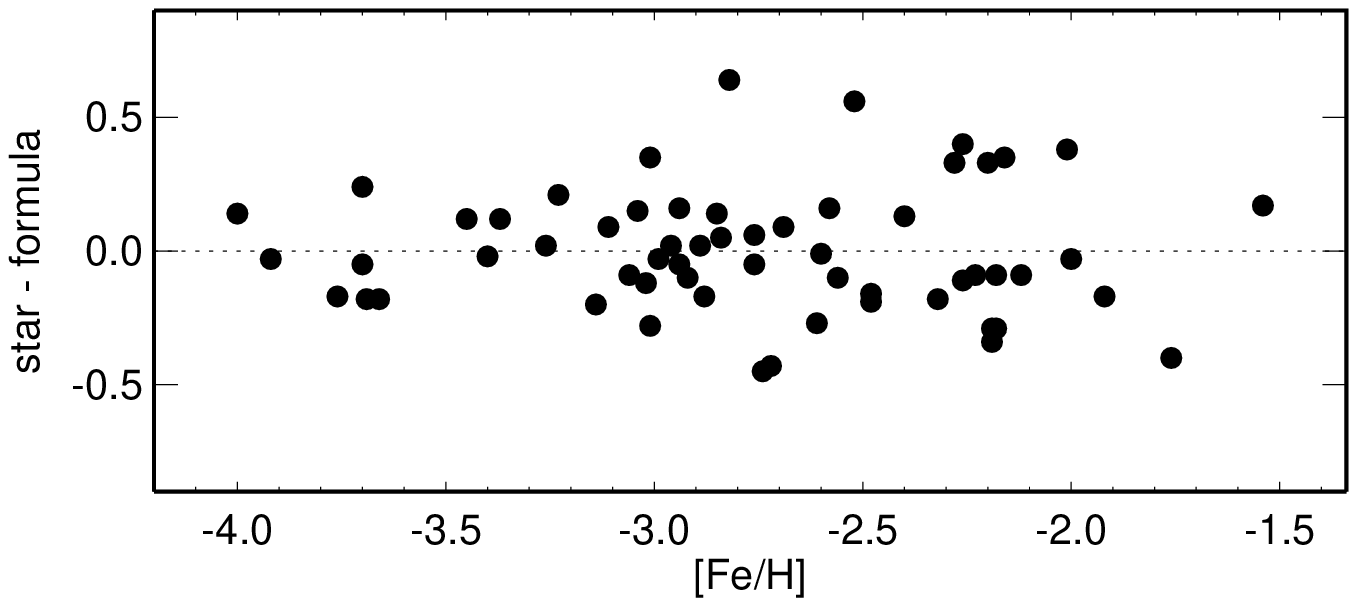}}

\vspace{-5mm}
\hspace{-3mm}  \resizebox{95mm}{!}{\includegraphics{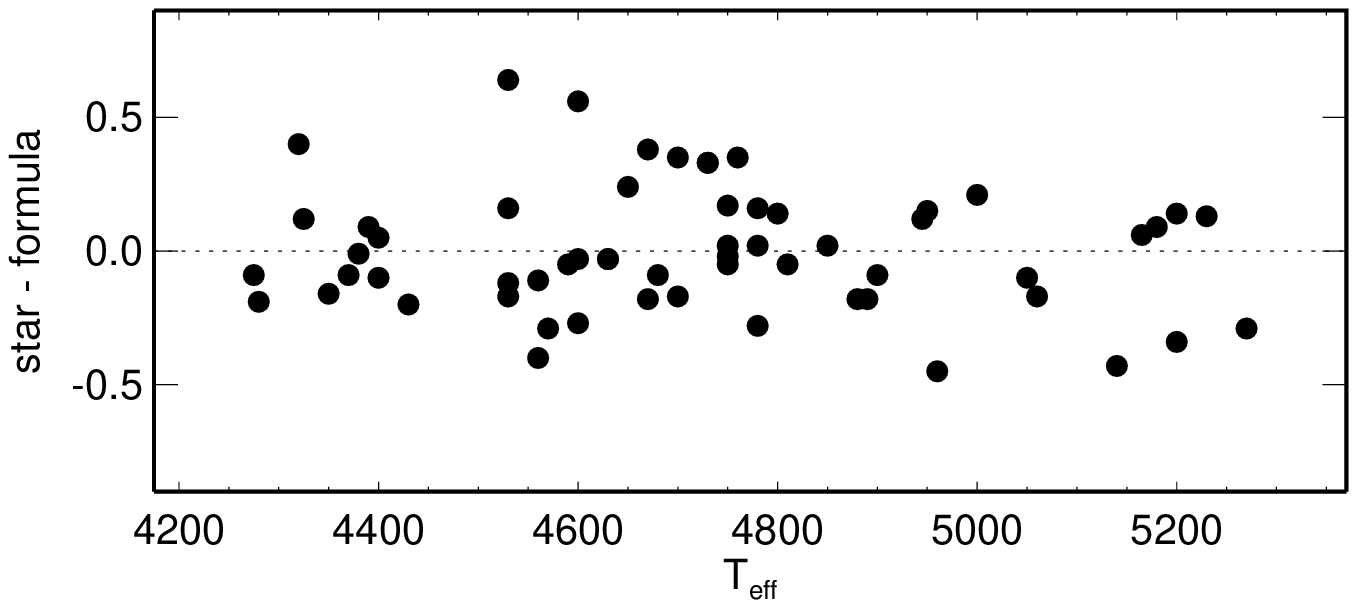}}
  \caption{\label{Fig:xi} Differences between the microturbulence velocity determined in this study for individual stars and that calculated with formula~(\ref{formula1}) as a function of metallicity (top panel) and effective temperature (bottom panel). }
\end{figure}

 Figure\,\ref{Fig:xi} indicates the deviation from the analytical fit
   of the individual determinations. The largest
  discrepancy of 0.56\,\kms\ is found for UMi-446, which also has the largest
  scatter of the \ion{Fe}{i} based abundances, with $\sigma_{\rm FeI}$ =
  0.27~dex. 

We tested the validity of Eq.\ref{formula1} on two ultra metal-poor (UMP) stars,
HE~0107-5240 (5100 / 2.2 / $-5.3$) and HE~0557-4840 (4900 / 2.2 / $-4.8$). Our
analytical fit gives $\xi_t$ = 2.0\,\kms\ and 1.9\,\kms, very close to the
determinations of \citet[][2.2\,\kms]{HE0107_ApJ} and
\citet[][1.8\,\kms]{Norrisetal:2007}, respectively.  Hence, we can only recommend to
use Eq.\ref{formula1} to calculate microturbulence velocities of EMP and UMP
giants.

\section{Comparison with other studies}\label{Sect:comparisons}

\begin{figure} 
\hspace{-3mm}  \resizebox{95mm}{!}{\includegraphics{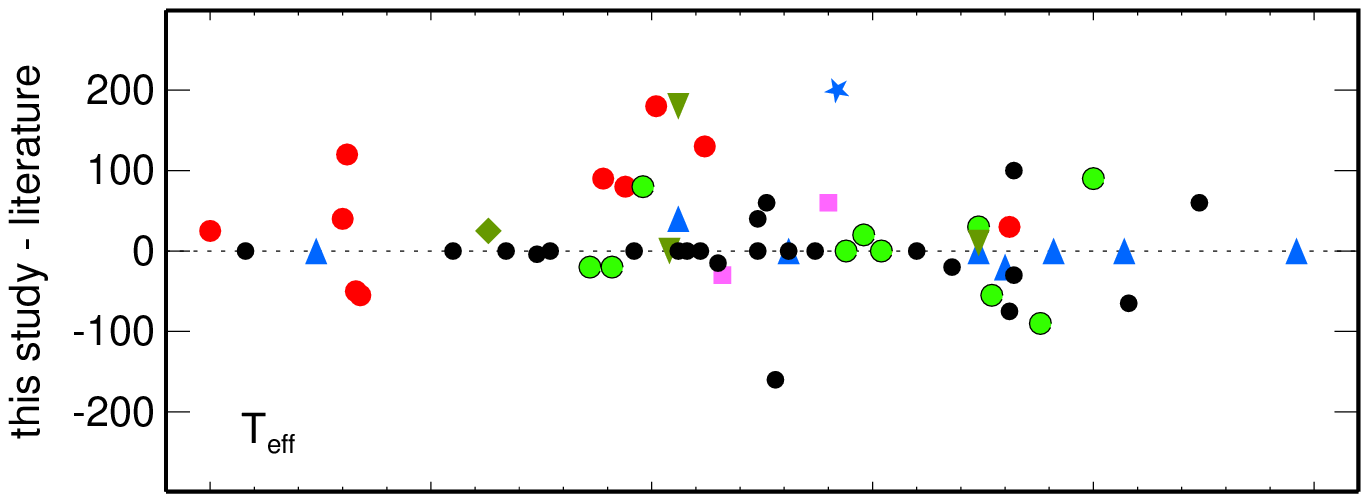}}

\vspace{-13mm}
\hspace{-3mm}  \resizebox{95mm}{!}{\includegraphics{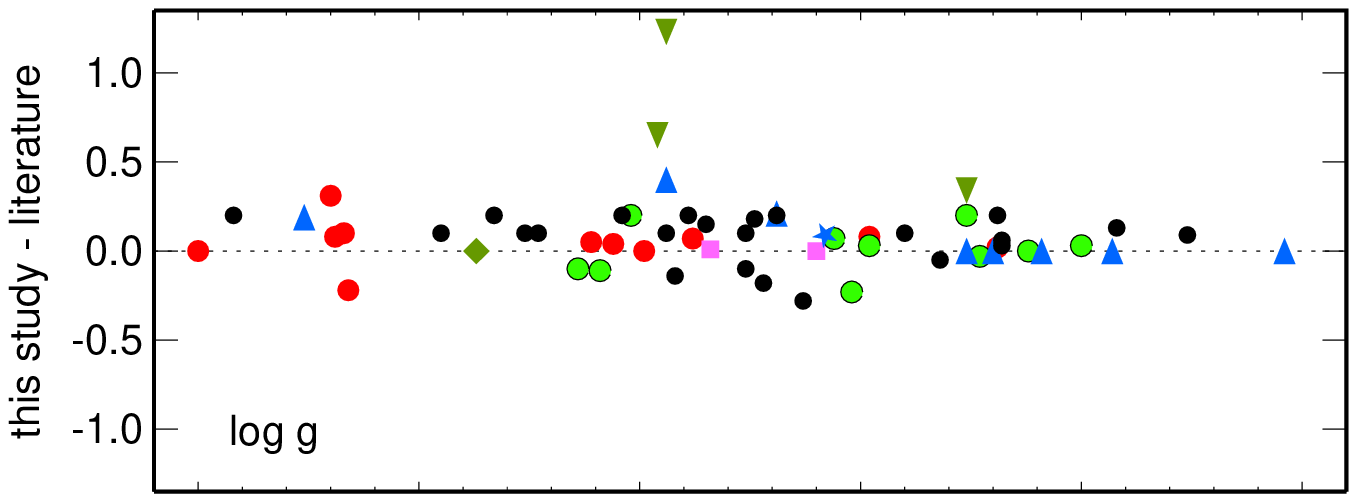}}

\vspace{-13mm}
\hspace{-3mm}  \resizebox{95mm}{!}{\includegraphics{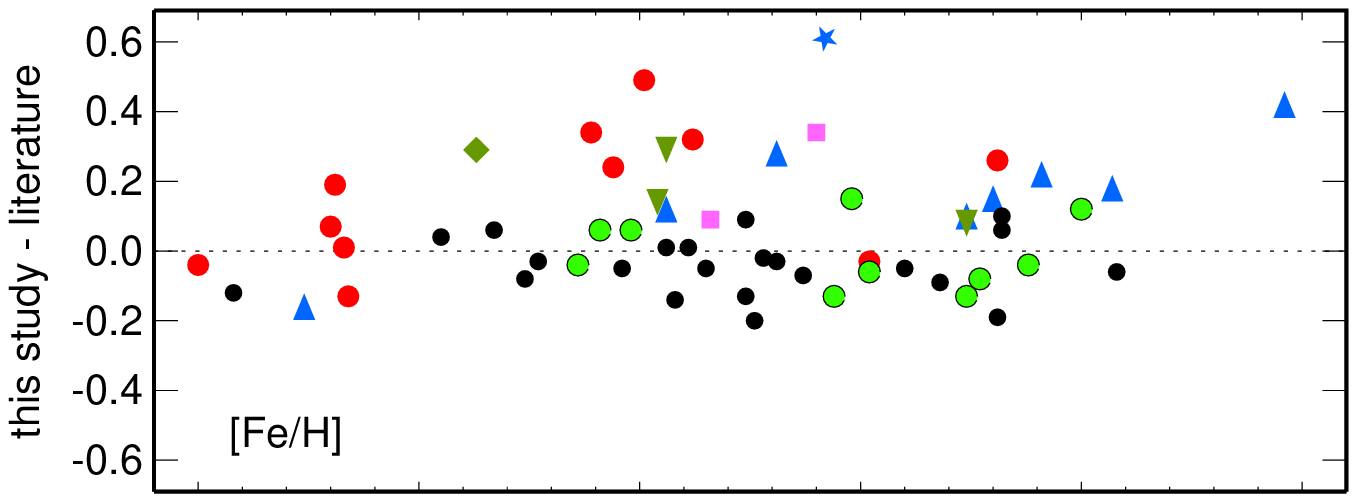}}

\vspace{-13mm}
\hspace{-3mm}  \resizebox{95mm}{!}{\includegraphics{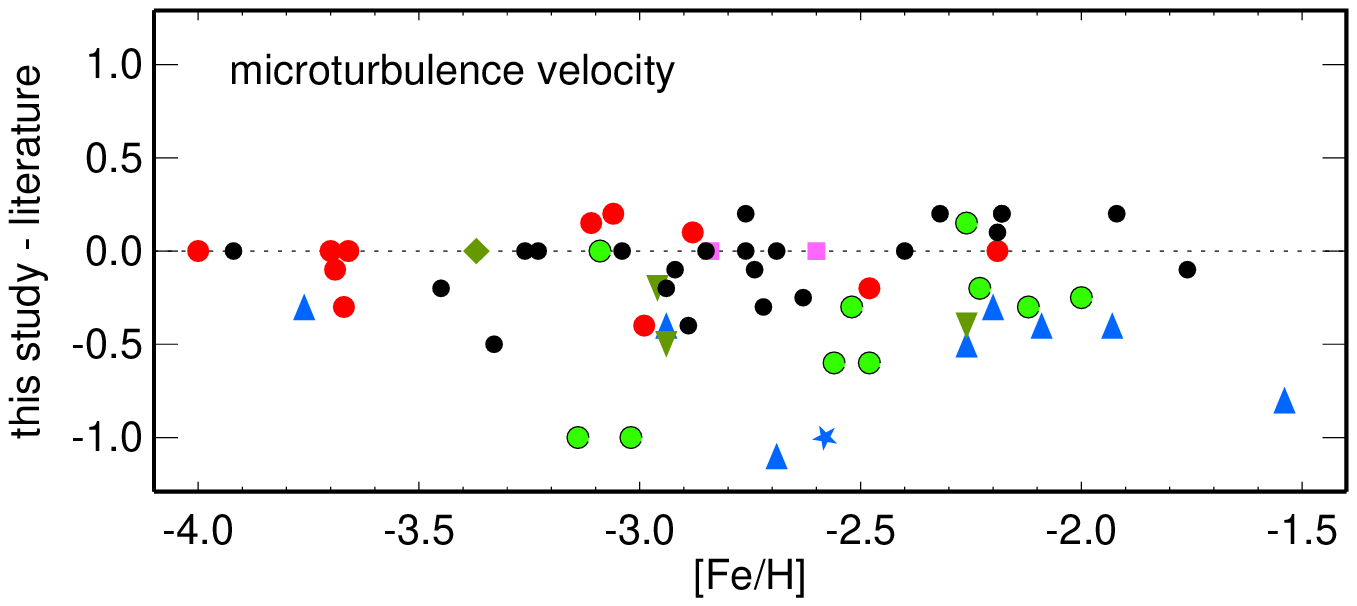}}
  \caption{\label{Fig:dsph_diff} Differences in atmospheric parameters, namely, $\Teff$, log~g, [Fe/H], and $\xi_t$ (in \kms), of the investigated stars between this and other studies. See text for references. Symbols as in Fig.~\ref{Fig:dnlte}. }
\end{figure}

The references to the different works from which our sample was built are listed
in Sect. \ref{Sect:basics}. While we produced a homogeneous set of atmospheric
parameters for the dSphs and MW populations that has no counterpart, it is
interesting to look back and identify the  origin of the changes. Not
all parameters have been impacted in the same way. For each star, Figure
\ref{Fig:dsph_diff} compares $\Teff$, $\xi_t$, log~g, and [Fe/H] in this study
and their previously published values.

The parameters of the Milky Way sample have hardly been modified
  at the exception of one or two stars. In contrast, the dSph sample has been
  notably impacted by the revision of the stellar atmospheric parameters and the
  NLTE treatment. These changes clearly depend on the original technique of analysis.

\subsection{log~g}

The surface gravity is the atmospheric parameter which has changed the
least.

We find that, for the dSphs, our final log~g values agree with the published
ones.  This is likely a consequence of the fact that some studies used the
distance-based log~g as an initial estimate of the stellar surface gravity,
which was then revised spectroscopically
\citep{2010ApJ...719..931C,2015A&A...583A..67J,2012AJ....144..168K,2010A&A...524A..58T}. Others,
such as \citet{Gilmore2013,Norris2010}, determined log~g from the isochrone
method. Both methods are close to our methodology.

There is one exception to this general agreement though. Surprisingly large
discrepancies of 0.65~dex and 1.23~dex between the distance based (this study)
and the LTE spectroscopic surface gravities \citep{2010ApJ...708..560F} were found for UMa~II-S1 and UMa~II-S2. They are much larger than the errors of log~g$_d$, 0.07~dex and 0.05dex, respectively, (Table\,\ref{Tab:parameters}) and cannot arise only due to the
LTE assumption because, at given atmospheric parameters, the NLTE abundance corrections for lines of \ion{Fe}{i}
are of the order of 0.07-0.08~dex, which are propagated to no more than 0.2~dex for an error of the log~g$_{sp}$.

For the MW giant sample, the literature data on log~g were mostly
obtained from non-spectroscopic methods, i.e., either the star's absolute
magnitude or / and its implied position in the color-magnitude diagram or / and an
average $\Teff$ versus log~g relationship for MP giants. The differences between
these data and our NLTE spectroscopic determinations mostly do not exceed
0.2~dex in absolute value.

\subsection{$\Teff$}

We have mentioned that this study is partly based on published photometric
temperatures. This is the case of the Bo{\"o}tes~I stars \citep{Norris2010,
Gilmore2013} and \citet{Cohen2013} sample of MW stars. For the
rest of the sample, the difference between the published $\Teff$ and ours
hardly exceeds 100~K. The previously published temperatures obtained by a spectroscopic method are systematically lower compared with our photometric temperatures. Such an effect was already pointed out in the literature \citep[see, for example][]{2013ApJ...769...57F}.

\subsection{Metallicity}

The differences between the published [Fe/H] values and those of the present
study are large.  They can be caused by a number of combined effects:

\begin{itemize}
\item different treatment of line formation, i.e., NLTE in this study and LTE in all other papers 
and using lines of \ion{Fe}{i} to derive final iron abundance in most cited papers that led to underestimated [Fe/H]; in contrast, our data are consistent within 0.15~dex with the values published by \citet{Cohen2013} who employed lines of \ion{Fe}{ii}, as we do, 
\item differences in derived microturbulence velocity,
\item differences in the used atomic parameters, in particular, van der Waals damping constants.
\end{itemize}

In part, a change in the stellar final metallicity is related to a correction of the original parameters. For example, Leo~IV-S1's final [Fe/H] value is 0.6~dex higher
  than in \citet{2010ApJ...716..446S}, as a consequence of our higher effective
  temperature, by 200~K from the original estimate.
The Sculptor stars at [Fe/H] $\sim -3$ have seen their
  effective temperatures substantially revised. An upward shift of about 0.2~dex in [Fe/H] is also caused by using lines of \ion{Fe}{ii} in this study, but not \ion{Fe}{i} under the LTE assumption, as in the literature. In case of Sex\,11-04, which sees essentially no change in $\Teff$ / log~g, the \ion{Fe}{i}-based LTE abundance of \citet{2010A&A...524A..58T} is 0.34~dex lower than our determination from lines of \ion{Fe}{ii}.

For Boo-041 we obtained an 0.42~dex higher iron abundance than that of \citet{Gilmore2013}, despite
  similar $\Teff$ / log~g = 4750~K / 1.6. This cannot be due to accounting for the
  NLTE effects, because (NLTE~-~LTE) = 0.06~dex for \ion{Fe}{i}. A difference of
  0.25~dex in $\eps{FeI}$ appears already in
  LTE, and this is due to a 0.8\,\kms\ lower microturbulence velocity in our
  study.  Going back to the LTE calculations and adopting the atmospheric parameters
  4750 / 1.6 / $-1.96$ and $\xi_t$ = 2.8\,\kms\ of \citet{Gilmore2013} result in a
  steep negative slope of $-0.41$ for the $\eps{FeI}$ - log~$W_{obs} / \lambda$
  plot, using 35 lines of \ion{Fe}{i} with \Eexc\ $>$ 1.2~eV and $W_{obs} <$
  180~m\AA. In addition, the abundance difference $\eps{FeI}$ - $\eps{FeII}$ =
  $-0.22$ is uncomfortably large. We derived $\xi_t$ = 2.0\,\kms\ by minimising
  the trend of the NLTE abundances of the \ion{Fe}{i} lines with $W_{obs}$. This makes also the
  NLTE abundances from the two ionisation stages of iron consistent within
  0.11~dex.

\subsection{Microturbulence}

We find that the microturbulence velocities derived in this study agree well
with the corresponding values of \citet{Cohen2013}, but they are lower, by up to
1.2\,\kms, compared with the data from most other papers. This explains mostly
positive differences in [Fe/H] between our and other studies. In case of the
Bo{\"o}tes~I stars, a source of the discrepancy in $\xi_t$ was fixed in a
private communication with David Yong. It appears to be connected with applying
outdated van der Waals damping constants in analysis of \citet{Gilmore2013}.

We found a metallicity-dependent discrepancy in $\xi_t$ between
\citet{2010ApJ...719..931C} and this study, from +0.3\,\kms\ at [Fe/H] $\simeq
-2$ up to +1.0\,\kms\ at [Fe/H] $\simeq -3$. An overestimation of
microturbulence velocity by \citet{2010ApJ...719..931C} was, most probably,
caused by treating Rayleigh scattering as LTE absorption.

  
\section{Conclusions and recommendations}\label{Sect:Conclusions}

This paper presents a homogeneous set of accurate atmospheric parameters for
a complete sample of 36 VMP and EMP stars in the classical dSphs in Sculptor,
Ursa Minor, Sextans, and Fornax and the UFDs Bo\"otes~I, UMa~II, and Leo~IV.
For the purpose of comparison between the Milky Way halo and satellite
populations in a companion paper, which presents the NLTE abundances of nine
chemical elements, from Na to Ba, we also derived atmospheric parameters of
23 VMP and EMP cool giants in the MW.

$\bullet$ Using the dSph stars with non-spectroscopic $\Teff$ / log~g parameters, 
we showed that the two
ionisation stages, \ion{Fe}{i} and \ion{Fe}{ii}, have consistent NLTE abundances,
 when the inelastic collisions with \ion{H}{i} are
treated with a scaling factor of \kH\ = 0.5 to the classic Drawinian rates. This
 justifies the \ion{Fe}{i}/\ion{Fe}{ii} ionisation equilibrium method to
determine surface gravity for VMP giants with unknown distances. 
The statistical error of log~g$_{sp}$ is estimated to be 0.2-0.3~dex, if the \ion{Fe}{i}~--~\ion{Fe}{ii} abundance difference is determined with an accuracy of 0.1~dex and better. The systematic error due to the uncertainty in our 1D-NLTE(\kH\ = 0.5) model is estimated to be $-0.1$~dex to $-0.23$~dex depending on stellar atmosphere parameters.
We caution against applying this method to the [Fe/H] $\precsim -3.7$ stars. For our four most metal-poor stars, 1D-NLTE fails to achieve the \ion{Fe}{i}/\ion{Fe}{ii} ionisation equilibrium.

$\bullet$ For each star the final atmospheric parameters were checked with the
\ion{Ti}{i}/\ion{Ti}{ii} ionisation equilibrium. No
imbalance was found except for four most metal-poor stars at [Fe/H]
$\le -3.5$. We suspect that this problem is linked to uncertainty in the determination of $\Teff$
at these very low metallicities.

$\bullet$ As a sanity check, we computed the distance-based log~g for the five stars with
available Gaia parallax measurements (Gaia Data Release~1). For three of them,
log~g$_{sp}$ is consistent within the error bars with log~g$_{Gaia}$. However,
there is one exception to this general agreement, HD~8724, with log~g$_{sp}$ -
log~g$_{Gaia}$ = $-0.76$. An inspection of the star's position in the log~g -
$\Teff$ plane also does not support its log~g$_{Gaia}$ = 2.05. It is evident,
measured parallax of HD~8724 needs to be double checked.

$\bullet$ The accuracy of the derived atmospheric parameters allowed us to derive an
analytical relation to calculate $\xi_t$ from $\Teff$, log~g, and [Fe/H].

$\bullet$ The lessons taken from this work lead us to spell a few recommendations to
accurately determine the atmospheric parameters of VMP and EMP giants:

\begin{itemize}
\item Derive the effective temperature from photometric methods.
\item Get the surface gravities from the star distances, when they are
  available. If not, the NLTE \ion{Fe}{i}/\ion{Fe}{ii} ionisation equilibrium has proven to be
  a robust alternative at [Fe/H] $\succsim -3.7$. We caution that, at low metallicity, LTE leads to
  underestimate log~g by up to 0.3~dex.
\item Calculate the metallicity from the \ion{Fe}{ii} lines, because they are only weakly sensitive to $\Teff$ variation and nearly free of the NLTE effects. Our study shows that the \ion{Fe}{i} lines under the LTE assumption lead to underestimate the stellar metallicity by up to 0.3~dex.
\item Check  $\Teff$ and log~g with theoretical evolutionary tracks.
\end{itemize}

\begin{acknowledgements}
  
This study is based on observations made with ESO Telescopes at the La Silla Paranal Observatory under programme IDs 68.D-0546(A), 71.B-0529(A), 076.D-0546(A), 079.B-0672A, 081.B-0620A, 087.D-0928A, 091.D-0912A,281.B-50220A, P82.182.B-0372, and P383.B-0038 and with the Canada-France-Hawaii Telescope under programme IDs 12BS04 and 05AC23. This research used the services of the ESO Science Archive Facility and the facilities of the Canadian Astronomy Data Centre operated by the National Research Council of Canada with the support of the Canadian Space Agency. We thank Judith G. Cohen, Rana Ezzeddine, Anna Frebel, and Joshua D. Simon for providing stellar spectra, Jelte de Jong for photometric data for Leo~IV-S1, and Paul Barklem for the \ion{Fe}{i}+\ion{H}{i} collision rate coefficients. 
L.M., Y.P., and T.S. are supported by the Presidium RAS
Programme P-7.  P.J and P.N acknowledge financial support from the Swiss National Science Foundation. T.S. acknowledges financial support from the Russian Foundation for Basic Research (grant 16-32-00695).
The authors are indebted to the International Space Science Institute (ISSI),
Bern, Switzerland, for supporting and funding the international teams “First
stars in dwarf spheroidal galaxies“ and "The Formation and Evolution of the
Galactic Halo". We made use the MARCS, SIMBAD, and VALD databases.

\end{acknowledgements}

\bibliography{mashonkina,atomic_data,nlte,scl,references,mp_stars}
\bibliographystyle{aa}

\begin{appendix}

\section{Line data}

\longtab[1]{
\begin{longtable}[]{lccrrclccrr}   
\caption{\label{Tab:linelist} Line data. $\Gamma_6$ corresponds to 10\,000~K.}  \\
\hline \noalign{\smallskip}
 Atom & $\lambda$ & $E_{\rm exc}$ & $\log gf$ & $\log \Gamma_6 / N_{\rm H}$ & \ \ \ & Atom & $\lambda$ & $E_{\rm exc}$ & $\log gf$ & $\log \Gamma_6 / N_{\rm H}$ \\  
      & (\AA)     & (eV)          &           &  (rad / s$\cdot$cm$^3$)     & &      & (\AA)     & (eV)          &           &  (rad / s$\cdot$cm$^3$) \\
 \noalign{\smallskip}  \hline \noalign{\smallskip} 
\endfirsthead
\caption{continued.}\\
\hline \noalign{\smallskip}
 Atom & $\lambda$ & $E_{\rm exc}$ & $\log gf$ & $\log \Gamma_6 / N_{\rm H}$ & \ \ \ & Atom & $\lambda$ & $E_{\rm exc}$ & $\log gf$ & $\log \Gamma_6 / N_{\rm H}$ \\  
      & (\AA)     & (eV)          &           &  (rad / s$\cdot$cm$^3$)     & &      & (\AA)     & (eV)          &           &  (rad / s$\cdot$cm$^3$) \\
 \noalign{\smallskip}  \hline  \noalign{\smallskip}    
\endhead
\hline
\endfoot
\hline
\endlastfoot
Ti I  & 3998.64 & 0.05 &  0.02 & $-$7.654   & & Fe I  & 4383.55 & 1.48 &  0.20   & $-$7.669 \\ 
Ti I  & 4533.25 & 0.85 &  0.54 & $-$7.626   & & Fe I  & 4404.75 & 1.56 & $-$0.14 & $-$7.659 \\ 
Ti I  & 4534.78 & 0.84 &  0.35 & $-$7.626   & & Fe I  & 4415.12 & 1.61 & $-$0.61 & $-$7.652 \\ 
Ti I  & 4548.77 & 0.83 & $-$0.28 & $-$7.626 & & Fe I  & 4430.61 & 2.22 & $-$1.66 & $-$7.511 \\ 
Ti I  & 4555.49 & 0.85 & $-$0.40 & $-$7.626 & & Fe I  & 4442.34 & 2.20 & $-$1.25 & $-$7.518 \\ 
Ti I  & 4840.87 & 0.90 & $-$0.43 & $-$7.697 & & Fe I  & 4443.19 & 2.86 & $-$1.04 & $-$7.788 \\ 
Ti I  & 4981.73 & 0.85 &  0.57 & $-$7.626   & & Fe I  & 4447.72 & 2.22 & $-$1.34 & $-$7.513 \\ 
Ti I  & 4991.06 & 0.84 &  0.45 & $-$7.629   & & Fe I  & 4459.12 & 2.18 & $-$1.28 & $-$7.525 \\ 
Ti I  & 4999.50 & 0.83 &  0.32 & $-$7.632   & & Fe I  & 4494.56 & 2.20 & $-$1.14 & $-$7.526 \\ 
Ti I  & 5014.28 & 0.81 &  0.04 & $-$7.635   & & Fe I  & 4531.15 & 1.49 & $-$2.15 & $-$7.790 \\ 
Ti I  & 5016.16 & 0.85 & $-$0.48 & $-$7.629 & & Fe I  & 4871.32 & 2.87 & $-$0.36 & $-$7.259 \\ 
Ti I  & 5039.96 & 0.02 & $-$1.08 & $-$7.720 & & Fe I  & 4872.14 & 2.88 & $-$0.57 & $-$7.255 \\ 
Ti I  & 5064.65 & 0.05 & $-$0.94 & $-$7.719 & & Fe I  & 4891.49 & 2.85 & $-$0.11 & $-$7.264 \\ 
Ti I  & 5173.74 & 0.00 & $-$1.06 & $-$7.729 & & Fe I  & 4903.31 & 2.88 & $-$0.93 & $-$7.259 \\
Ti I  & 5192.97 & 0.02 & $-$0.95 & $-$7.727 & & Fe I  & 4918.99 & 2.87 & $-$0.34 & $-$7.264 \\
Ti I  & 5210.39 & 0.05 & $-$0.82 & $-$7.724 & & Fe I  & 4920.50 & 2.83 &  0.07   & $-$7.271 \\ 
Ti II & 3913.47 & 1.12 & $-$0.36 & $-$7.896 & & Fe I  & 4938.81 & 2.87 & $-$1.08 & $-$7.264 \\ 
Ti II & 4012.39 & 0.57 & $-$1.78 & $-$7.909 & & Fe I  & 4966.10 & 3.33 & $-$0.89 & $-$7.218 \\
Ti II & 4028.34 & 1.89 & $-$0.92 & $-$7.908 & & Fe I  & 5001.86 & 3.88 &  0.01   & $-$7.273 \\
Ti II & 4290.22 & 1.16 & $-$0.87 & $-$7.915 & & Fe I  & 5006.12 & 2.83 & $-$0.63 & $-$7.280 \\
Ti II & 4300.05 & 1.18 & $-$0.46 & $-$7.909 & & Fe I  & 5041.76 & 1.49 & $-$2.20 & $-$7.810 \\
Ti II & 4337.92 & 1.08 & $-$0.96 & $-$7.923 & & Fe I  & 5049.82 & 2.28 & $-$1.36 & $-$7.586 \\
Ti II & 4394.05 & 1.22 & $-$1.77 & $-$7.944 & & Fe I  & 5068.77 & 2.94 & $-$1.04 & $-$7.265 \\
Ti II & 4395.03 & 1.08 & $-$0.54 & $-$7.920 & & Fe I  & 5074.75 & 4.22 & $-$0.20 & $-$7.189 \\
Ti II & 4395.85 & 1.24 & $-$1.93 & $-$7.904 & & Fe I  & 5159.05 & 4.28 & $-$0.81 & $-$7.175 \\
Ti II & 4399.77 & 1.24 & $-$1.20 & $-$7.946 & & Fe I  & 5162.29 & 4.18 &  0.02   & $-$7.239 \\
Ti II & 4417.72 & 1.16 & $-$1.19 & $-$7.926 & & Fe I  & 5171.61 & 1.48 & $-$1.75 & $-$7.687 \\
Ti II & 4418.33 & 1.24 & $-$1.99 & $-$7.840 & & Fe I  & 5191.45 & 3.04 & $-$0.55 & $-$7.258 \\
Ti II & 4443.79 & 1.08 & $-$0.71 & $-$7.923 & & Fe I  & 5192.34 & 3.00 & $-$0.52 & $-$7.266 \\
Ti II & 4444.56 & 1.12 & $-$2.20 & $-$7.931 & & Fe I  & 5194.94 & 1.56 & $-$2.09 & $-$7.680 \\
Ti II & 4450.48 & 1.08 & $-$1.52 & $-$7.920 & & Fe I  & 5215.19 & 3.27 & $-$0.93 & $-$7.203 \\
Ti II & 4464.45 & 1.16 & $-$1.81 & $-$7.926 & & Fe I  & 5216.28 & 1.61 & $-$2.10 & $-$7.674 \\
Ti II & 4468.51 & 1.13 & $-$0.63 & $-$7.931 & & Fe I  & 5232.95 & 2.94 & $-$0.07 & $-$7.280 \\
Ti II & 4470.86 & 1.16 & $-$2.02 & $-$7.928 & & Fe I  & 5266.56 & 3.00 & $-$0.39 & $-$7.273 \\
Ti II & 4501.27 & 1.12 & $-$0.77 & $-$7.851 & & Fe I  & 5281.79 & 3.04 & $-$0.83 & $-$7.266 \\
Ti II & 4533.96 & 1.24 & $-$0.53 & $-$7.960 & & Fe I  & 5283.62 & 3.24 & $-$0.52 & $-$7.221 \\
Ti II & 4563.76 & 1.22 & $-$0.69 & $-$7.961 & & Fe I  & 5302.30 & 3.28 & $-$0.88 & $-$7.210 \\
Ti II & 4571.97 & 1.57 & $-$0.31 & $-$7.894 & & Fe I  & 5307.37 & 1.61 & $-$2.99 & $-$7.678 \\
Ti II & 4583.41 & 1.16 & $-$2.84 & $-$7.928 & & Fe I  & 5324.19 & 3.21 & $-$0.10 & $-$7.235 \\
Ti II & 4657.20 & 1.24 & $-$2.29 & $-$7.850 & & Fe I  & 5328.53 & 1.56 & $-$1.85 & $-$7.686 \\
Ti II & 4708.67 & 1.24 & $-$2.35 & $-$7.850 & & Fe I  & 5339.93 & 3.27 & $-$0.68 & $-$7.221 \\
Ti II & 4798.53 & 1.08 & $-$2.66 & $-$7.923 & & Fe I  & 5364.86 & 4.45 &  0.22   & $-$7.136 \\
Ti II & 4865.61 & 1.12 & $-$2.70 & $-$7.950 & & Fe I  & 5367.48 & 4.42 &  0.55   & $-$7.153 \\
Ti II & 5129.16 & 1.89 & $-$1.34 & $-$7.908 & & Fe I  & 5369.96 & 4.37 &  0.54   & $-$7.179 \\
Ti II & 5154.07 & 1.57 & $-$1.75 & $-$7.950 & & Fe I  & 5383.37 & 4.31 &  0.50   & $-$7.219 \\
Ti II & 5185.91 & 1.89 & $-$1.41 & $-$7.908 & & Fe I  & 5389.48 & 4.42 & $-$0.40 & $-$7.159 \\
Ti II & 5188.68 & 1.58 & $-$1.05 & $-$7.948 & & Fe I  & 5393.17 & 3.24 & $-$0.71 & $-$7.235 \\
Ti II & 5226.55 & 1.57 & $-$1.26 & $-$7.953 & & Fe I  & 5400.51 & 4.37 & $-$0.15 & $-$7.187 \\
Ti II & 5336.77 & 1.58 & $-$1.60 & $-$7.953 & & Fe I  & 5415.19 & 4.39 &  0.51   & $-$7.182 \\
Ti II & 5381.01 & 1.57 & $-$1.97 & $-$7.956 & & Fe I  & 5424.07 & 4.32 &  0.52   & $-$7.224 \\
Ti II & 5418.77 & 1.58 & $-$2.13 & $-$7.953 & & Fe I  & 5569.62 & 3.42 & $-$0.54 & $-$7.204 \\
Fe I  & 3753.61 & 2.18 & $-$0.89 & $-$7.815 & & Fe I  & 5572.84 & 3.40 & $-$0.31 & $-$7.211 \\
Fe I  & 3765.54 & 3.24 &  0.48   & $-$7.790 & & Fe I  & 5576.09 & 3.43 & $-$1.00 & $-$7.201 \\
Fe I  & 3805.34 & 3.30 &  0.31   & $-$7.683 & & Fe I  & 5586.76 & 3.37 & $-$0.14 & $-$7.221 \\
Fe I  & 3815.84 & 1.48 &  0.24   & $-$7.608 & & Fe I  & 5615.66 & 3.33 &  0.05   & $-$7.234 \\
Fe I  & 3827.82 & 1.56 &  0.06   & $-$7.597 & & Fe I  & 6003.03 & 3.88 & $-$1.11 & $-$7.181 \\
Fe I  & 3997.39 & 2.73 & $-$0.40 & $-$7.757 & & Fe I  & 6024.05 & 4.55 & $-$0.11 & $-$7.225 \\
Fe I  & 4005.24 & 1.56 & $-$0.61 & $-$7.620 & & Fe I  & 6136.62 & 2.45 & $-$1.50 & $-$7.609 \\
Fe I  & 4021.87 & 2.76 & $-$0.66 & $-$7.755 & & Fe I  & 6137.70 & 2.59 & $-$1.37 & $-$7.589 \\
Fe I  & 4032.63 & 1.48 & $-$2.44 & $-$7.599 & & Fe I  & 6173.34 & 2.22 & $-$2.85 & $-$7.690 \\
Fe I  & 4045.81 & 1.48 &  0.28   & $-$7.638 & & Fe I  & 6191.57 & 2.43 & $-$1.42 & $-$7.615 \\
Fe I  & 4063.59 & 1.56 &  0.07   & $-$7.627 & & Fe I  & 6200.31 & 2.61 & $-$2.44 & $-$7.589 \\
Fe I  & 4067.98 & 3.21 & $-$0.42 & $-$7.270 & & Fe I  & 6213.43 & 2.22 & $-$2.48 & $-$7.691 \\
Fe I  & 4071.74 & 1.61 & $-$0.02 & $-$7.619 & & Fe I  & 6219.29 & 2.20 & $-$2.44 & $-$7.694 \\
Fe I  & 4107.49 & 2.83 & $-$0.72 & $-$7.659 & & Fe I  & 6230.74 & 2.56 & $-$1.28 & $-$7.597 \\
Fe I  & 4132.06 & 1.61 & $-$0.67 & $-$7.626 & & Fe I  & 6240.66 & 2.22 & $-$3.23 & $-$7.661 \\
Fe I  & 4132.90 & 2.84 & $-$0.92 & $-$7.659 & & Fe I  & 6252.57 & 2.40 & $-$1.76 & $-$7.621 \\
Fe I  & 4134.68 & 2.83 & $-$0.49 & $-$7.661 & & Fe I  & 6265.13 & 2.18 & $-$2.55 & $-$7.700 \\
Fe I  & 4137.00 & 3.41 & $-$0.55 & $-$7.665 & & Fe I  & 6297.80 & 2.22 & $-$2.74 & $-$7.694 \\
Fe I  & 4143.87 & 1.56 & $-$0.46 & $-$7.636 & & Fe I  & 6301.50 & 3.65 & $-$0.72 & $-$7.540 \\
Fe I  & 4147.67 & 1.48 & $-$2.10 & $-$7.648 & & Fe I  & 6302.49 & 3.69 & $-$1.15 & $-$7.540 \\
Fe I  & 4154.50 & 2.83 & $-$0.69 & $-$7.760 & & Fe I  & 6322.69 & 2.59 & $-$2.43 & $-$7.596 \\
Fe I  & 4154.81 & 3.37 & $-$0.37 & $-$7.229 & & Fe I  & 6335.33 & 2.20 & $-$2.23 & $-$7.698 \\
Fe I  & 4156.80 & 2.83 & $-$0.81 & $-$7.663 & & Fe I  & 6344.15 & 2.43 & $-$2.92 & $-$7.620 \\
Fe I  & 4157.78 & 3.42 & $-$0.40 & $-$7.500 & & Fe I  & 6355.04 & 2.84 & $-$2.29 & $-$7.599 \\
Fe I  & 4175.64 & 2.84 & $-$0.68 & $-$7.663 & & Fe I  & 6393.61 & 2.43 & $-$1.43 & $-$7.622 \\
Fe I  & 4176.57 & 3.36 & $-$0.62 & $-$7.510 & & Fe I  & 6400.00 & 3.60 & $-$0.52 & $-$7.232 \\
Fe I  & 4181.76 & 2.83 & $-$0.37 & $-$7.665 & & Fe I  & 6408.03 & 3.69 & $-$1.00 & $-$7.540 \\
Fe I  & 4182.38 & 3.02 & $-$1.19 & $-$7.811 & & Fe I  & 6421.36 & 2.28 & $-$2.01 & $-$7.620 \\
Fe I  & 4184.89 & 2.83 & $-$0.84 & $-$7.780 & & Fe I  & 6430.86 & 2.18 & $-$1.95 & $-$7.704 \\
Fe I  & 4187.04 & 2.45 & $-$0.55 & $-$7.252 & & Fe I  & 6494.98 & 2.40 & $-$1.27 & $-$7.629 \\
Fe I  & 4187.80 & 2.42 & $-$0.55 & $-$7.258 & & Fe I  & 6593.88 & 2.43 & $-$2.39 & $-$7.629 \\
Fe I  & 4191.43 & 2.47 & $-$0.73 & $-$7.249 & & Fe I  & 6609.12 & 2.56 & $-$2.66 & $-$7.610 \\
Fe I  & 4195.33 & 3.33 & $-$0.41 & $-$7.540 & & Fe II & 4923.92 & 2.89 & $-$1.39 & $-$7.884 \\
Fe I  & 4199.10 & 3.05 &  0.25   & $-$7.678 & & Fe II & 5018.43 & 2.89 & $-$1.23 & $-$7.886 \\
Fe I  & 4202.03 & 1.48 & $-$0.70 & $-$7.653 & & Fe II & 5197.57 & 3.23 & $-$2.24 & $-$7.880 \\
Fe I  & 4213.65 & 2.84 & $-$1.30 & $-$7.780 & & Fe II & 5234.63 & 3.22 & $-$2.17 & $-$7.880 \\
Fe I  & 4222.21 & 2.45 & $-$0.97 & $-$7.258 & & Fe II & 5264.81 & 3.23 & $-$3.02 & $-$7.875 \\
Fe I  & 4227.43 & 3.33 &  0.23   & $-$7.550 & & Fe II & 5276.00 & 3.20 & $-$2.10 & $-$7.883 \\
Fe I  & 4233.60 & 2.48 & $-$0.60 & $-$7.252 & & Fe II & 5284.10 & 2.89 & $-$3.09 & $-$7.887 \\
Fe I  & 4238.81 & 3.40 & $-$0.27 & $-$7.243 & & Fe II & 5325.56 & 3.22 & $-$3.21 & $-$7.887 \\
Fe I  & 4250.12 & 2.47 & $-$0.40 & $-$7.258 & & Fe II & 5414.08 & 3.22 & $-$3.53 & $-$7.880 \\
Fe I  & 4260.47 & 2.40 & $-$0.02 & $-$7.274 & & Fe II & 5425.25 & 3.20 & $-$3.28 & $-$7.886 \\
Fe I  & 4271.15 & 2.45 & $-$0.35 & $-$7.266 & & Fe II & 5534.85 & 3.24 & $-$2.75 & $-$7.883 \\
Fe I  & 4271.76 & 1.48 & $-$0.16 & $-$7.660 & & Fe II & 6247.56 & 3.89 & $-$2.33 & $-$7.870 \\
Fe I  & 4282.40 & 2.17 & $-$0.82 & $-$7.830 & & Fe II & 6432.68 & 2.89 & $-$3.58 & $-$7.899 \\
Fe I  & 4325.76 & 1.61 & $-$0.01 & $-$7.645 & & Fe II & 6456.39 & 3.90 & $-$2.07 & $-$7.873 \\
Fe I  & 4337.05 & 1.56 & $-$1.70 & $-$7.654 & & Fe II & 6516.08 & 2.89 & $-$3.32 & $-$7.899 \\
Fe I  & 4352.73 & 2.22 & $-$1.26 & $-$7.830 & &       &         &      &         &          \\ 
\hline 
\end{longtable}
}

\longtab[2]{
\begin{longtable}[]{lcllll}   
\caption{\label{Tab:fe_ti} Iron and titanium NLTE abundances for the investigated sample. Numbers in parenthesis indicate $\sigma_{\eps{}}$ and the number of lines measured.}  \\
\hline \noalign{\smallskip}
\multicolumn{1}{c}{ID} & $\Teff$ / log~g / [Fe/H] / $\xi_t$ & \ion{Fe}{i} & \ion{Fe}{ii} & \ion{Ti}{i} & \ion{Ti}{ii} \\  
 \noalign{\smallskip}  \hline \noalign{\smallskip} 
\endfirsthead
\caption{continued.}\\
\hline \noalign{\smallskip}
\multicolumn{1}{c}{ID} & $\Teff$ / log~g / [Fe/H] / $\xi_t$ & \multicolumn{1}{c}{ \ion{Fe}{i}} & \multicolumn{1}{c}{ \ion{Fe}{ii}} & \multicolumn{1}{c}{ \ion{Ti}{i}} & \multicolumn{1}{c}{ \ion{Ti}{ii}} \\  
 \noalign{\smallskip}  \hline  \noalign{\smallskip}    
\endhead
\hline
\endfoot
\hline
\endlastfoot
Scl ET0381          & 4570 / 1.17 / $-2.19$ / 1.7 &  5.23 (0.17, 74) &  5.31 (0.09,  9) &  2.31 (0.15, 10) &  2.60 (0.17, 21) \\
Scl002\_06          & 4390 / 0.68 / $-3.11$ / 2.3 &  4.33 (0.16, 69) &  4.39 (0.09,  4) &  1.92 (0.16,  4) &  1.75 (0.08, 13) \\
Scl03\_059          & 4530 / 1.08 / $-2.88$ / 1.9 &  4.60 (0.16, 91) &  4.62 (0.10,  4) &  2.35 (0.12, 11) &  2.36 (0.18, 19) \\
Scl031\_11          & 4670 / 1.13 / $-3.69$ / 2.0 &  4.11 (0.14, 37) &  3.81 (0.07,  2) &  2.11 (0.13,  3) &  1.70 (0.18, 12) \\
Scl074\_02          & 4680 / 1.23 / $-3.06$ / 2.0 &  4.50 (0.21, 56) &  4.44 (0.12,  5) &  2.15 (0.20,  4) &  1.85 (0.23, 17) \\
Scl07-49            & 4630 / 1.28 / $-2.99$ / 2.0 &  4.64 (0.15, 22) &  4.59 (0.16,  4) &  2.35 (0.19,  6) &  2.31 (0.24,  3) \\
Scl07-50            & 4800 / 1.56 / $-4.00$ / 2.2 &  3.86 (0.13, 17) &  3.50 (0.20,  2) &  &  1.29 (0.13,  9) \\
Scl11\_1\_4296      & 4810 / 1.76 / $-3.70$ / 1.9 &  3.80 (0.42, 21) &  3.80 (0.00,  2) &  &  1.56 (0.16, 11) \\
Scl6\_6\_402        & 4890 / 1.78 / $-3.66$ / 1.8 &  4.16 (0.35, 20) &  3.84 (0.04,  4) &  &  1.68 (0.29,  4) \\
Scl S1020549        & 4650 / 1.35 / $-3.67$ / 2.0 &  3.99 (0.27, 38) &  3.83 (0.15,  5) &  2.01 (0.03,  2) &  1.64 (0.23, 16) \\
Scl1019417          & 4280 / 0.50 / $-2.48$ / 2.0 &  5.12 (0.16, 33) &  5.02 (0.18, 10) &  2.90 (0.13,  8) &  2.98 (0.09,  5) \\
Fnx05-42            & 4350 / 0.70 / $-3.37$ / 2.3 &  4.03 (0.17, 20) &  4.13 (0.08,  2) &  1.68 (0.06,  3) &  1.78 (0.04,  6) \\
Sex11-04            & 4380 / 0.57 / $-2.60$ / 2.2 &  4.84 (0.13, 37) &  4.90 (0.12,  4) &  2.49 (0.13,  7) &  2.42 (0.10,  7) \\
Sex24-72            & 4400 / 0.76 / $-2.84$ / 2.2 &  4.71 (0.14, 43) &  4.66 (0.05,  3) &  2.22 (0.10,  6) &  2.24 (0.10,  7) \\
UMi396              & 4320 / 0.70 / $-2.26$ / 2.5 &  5.13 (0.14, 26) &  5.24 (0.05,  5) &  3.10 (0.28,  8) &  2.92 (0.29, 10) \\
UMi446              & 4600 / 1.37 / $-2.52$ / 2.5 &  5.00 (0.27, 28) &  4.98 (0.28,  3) &  3.00 (0.27,  4) &  2.96 (0.44,  7) \\
UMi718              & 4630 / 1.13 / $-2.00$ / 2.0 &  5.56 (0.18, 32) &  5.50 (0.12,  3) &  3.16 (0.11,  6) &  3.15 (0.18,  6) \\
UMi COS233          & 4370 / 0.77 / $-2.23$ / 2.0 &  5.23 (0.11, 29) &  5.27 (0.11,  8) &  2.86 (0.14,  9) &  3.12 (0.21,  5) \\
UMi JI19            & 4530 / 1.00 / $-3.02$ / 2.0 &  4.50 (0.22, 42) &  4.48 (0.15,  9) &  2.47 (0.06,  4) &  2.33 (0.18,  5) \\
UMi20103            & 4780 / 1.55 / $-3.09$ / 2.0 &  4.53 (0.14, 34) &  4.41 (0.17,  7) &  2.20 (0.18,  2) &  1.91 (0.23,  4) \\
UMi28104            & 4275 / 0.65 / $-2.12$ / 2.0 &  5.23 (0.13, 25) &  5.38 (0.07,  6) &  2.73 (0.11,  9) &  3.06 (0.11,  6) \\
UMi33533            & 4430 / 0.75 / $-3.14$ / 2.0 &  4.33 (0.16, 42) &  4.36 (0.12, 10) &  1.96 (0.09,  6) &  1.91 (0.15,  9) \\
UMi36886            & 4400 / 0.82 / $-2.56$ / 2.0 &  4.92 (0.15, 40) &  4.94 (0.09,  8) &  2.52 (0.15,  9) &  2.79 (0.08,  6) \\
UMi41065            & 4350 / 0.63 / $-2.48$ / 2.0 &  5.05 (0.11, 34) &  5.02 (0.17,  7) &  2.67 (0.17,  9) &  2.88 (0.19,  6) \\
Boo-033             & 4730 / 1.40 / $-2.26$ / 2.3 &  5.29 (0.17, 33) &  5.24 (0.16,  4) &  2.73 (0.30,  5) &  2.74 (0.23,  3) \\
Boo-041             & 4750 / 1.60 / $-1.54$ / 2.0 &  5.86 (0.23, 16) &  5.96 (0.07,  2) &  4.17 (0.14,  2) &  4.19 (0.20,  1) \\
Boo-094             & 4570 / 1.01 / $-2.69$ / 2.2 &  4.72 (0.15, 41) &  4.81 (0.08,  2) &  2.44 (0.13,  6) &  2.35 (0.10,  5) \\
Boo-117             & 4700 / 1.40 / $-2.09$ / 2.3 &  5.38 (0.33, 35) &  5.41 (0.23,  3) &  2.93 (0.11,  6) &  2.86 (0.14,  4) \\
Boo-127             & 4670 / 1.40 / $-1.93$ / 2.3 &  5.49 (0.18, 20) &  5.57 (0.03,  3) &  3.20 (0.27,  7) &  3.09 (0.02,  3) \\
Boo-130             & 4730 / 1.40 / $-2.20$ / 2.3 &  5.26 (0.19, 33) &  5.30 (0.06,  2) &  2.72 (0.15,  3) &  2.92 (0.14,  3) \\
Boo-980             & 4760 / 1.80 / $-2.94$ / 1.8 &  4.59 (0.19, 49) &  4.56 (0.19,  9) &  2.57 (0.13,  5) &  2.52 (0.18, 23) \\
Boo-1137            & 4700 / 1.39 / $-3.76$ / 1.9 &  4.01 (0.16, 39) &  3.74 (0.06,  2) &  2.10 (0.10,  5) &  1.93 (0.15, 17) \\
UMa~II-S1           & 4850 / 2.05 / $-2.96$ / 1.8 &  4.47 (0.18, 35) &  4.54 (0.14,  7) &  2.27 (0.23,  3) &  2.11 (0.25,  9) \\
UMa~II-S2           & 4780 / 1.83 / $-2.94$ / 2.0 &  4.46 (0.15, 24) &  4.56 (0.18,  6) &  2.24 (0.19,  4) &  2.13 (0.39,  9) \\
UMa~II-S3           & 4560 / 1.34 / $-2.26$ / 1.8 &  5.16 (0.14, 45) &  5.24 (0.10, 11) &  2.74 (0.08, 12) &  2.80 (0.14, 11) \\
Leo~IV-S1           & 4530 / 1.09 / $-2.58$ / 2.2 &  4.80 (0.22, 32) &  4.92 (0.13,  4) &  2.43 (0.23,  2) &  2.27 (0.23,  9) \\
HD~2796             & 4880 / 1.55 / $-2.32$ / 1.8 &  5.13 (0.07, 45) &  5.18 (0.03,  6) &  2.89 (0.06, 11) &  2.83 (0.05, 14) \\
HD~4306             & 4960 / 2.18 / $-2.74$ / 1.3 &  4.71 (0.09, 45) &  4.76 (0.06,  6) &  2.63 (0.02,  8) &  2.59 (0.05, 11) \\
HD~8724             & 4560 / 1.29 / $-1.76$ / 1.5 &  5.70 (0.10, 32) &  5.74 (0.04,  5) &  3.30 (0.07,  8) &  3.38 (0.04, 12) \\
HD~108317           & 5270 / 2.96 / $-2.18$ / 1.2 &  5.27 (0.10, 45) &  5.32 (0.05,  6) &  3.03 (0.04,  9) &  3.05 (0.03,  8) \\
HD~122563           & 4600 / 1.32 / $-2.63$ / 1.7 &  4.82 (0.07, 39) &  4.87 (0.03,  4) &  2.48 (0.02,  9) &  2.54 (0.06, 13) \\
HD~128279           & 5200 / 3.00 / $-2.19$ / 1.1 &  5.29 (0.09, 45) &  5.31 (0.04,  6) &  3.01 (0.01,  8) &  3.05 (0.03, 13) \\
HD~218857           & 5060 / 2.53 / $-1.92$ / 1.4 &  5.55 (0.10, 45) &  5.58 (0.05,  6) &  3.25 (0.04,  8) &  3.28 (0.03, 10) \\
HE0011-0035         & 4950 / 2.00 / $-3.04$ / 2.0 &  4.44 (0.21, 38) &  4.46 (0.15, 11) &  2.41 (0.10,  5) &  2.32 (0.25, 14) \\
HE0039-4154         & 4780 / 1.60 / $-3.26$ / 2.0 &  4.30 (0.22, 41) &  4.24 (0.18,  7) &  1.91 (0.17,  4) &  1.81 (0.15, 12) \\
HE0048-0611         & 5180 / 2.70 / $-2.69$ / 1.7 &  4.84 (0.14, 44) &  4.81 (0.10, 12) &  2.72 (0.09,  8) &  2.71 (0.14, 16) \\
HE0122-1616         & 5200 / 2.65 / $-2.85$ / 1.8 &  4.65 (0.15, 41) &  4.65 (0.09, 11) &  2.39 (0.11,  5) &  2.23 (0.12, 10) \\
HE0332-1007         & 4750 / 1.50 / $-2.89$ / 2.0 &  4.59 (0.15, 41) &  4.61 (0.09, 12) &  2.31 (0.05,  5) &  2.38 (0.13,  9) \\
HE0445-2339         & 5165 / 2.20 / $-2.76$ / 1.9 &  4.74 (0.09, 42) &  4.74 (0.06, 13) &  2.58 (0.05,  9) &  2.43 (0.10, 18) \\
HE1356-0622         & 4945 / 2.00 / $-3.45$ / 2.0 &  4.01 (0.12, 35) &  4.05 (0.08,  8) &  2.03 (0.03,  4) &  1.75 (0.10, 12) \\
HE1357-0123         & 4600 / 1.20 / $-3.92$ / 2.1 &  3.76 (0.17, 36) &  3.58 (0.09,  6) &  1.60 (0.24,  3) &  1.28 (0.13, 11) \\
HE1416-1032         & 5000 / 2.00 / $-3.23$ / 2.1 &  4.25 (0.14, 41) &  4.27 (0.11,  8) &  2.05 (0.07,  3) &  1.89 (0.10, 11) \\
HE2244-2116         & 5230 / 2.80 / $-2.40$ / 1.7 &  5.12 (0.12, 42) &  5.10 (0.06, 12) &  2.97 (0.11,  9) &  2.99 (0.09, 18) \\
HE2249-1704         & 4590 / 1.20 / $-2.94$ / 2.0 &  4.58 (0.14, 28) &  4.56 (0.07, 12) &  2.22 (0.11, 10) &  2.27 (0.14, 13) \\
HE2252-4225         & 4750 / 1.55 / $-2.76$ / 1.9 &  4.73 (0.09, 32) &  4.74 (0.07,  8) &  2.54 (0.05,  8) &  2.58 (0.05, 13) \\
HE2327-5642         & 5050 / 2.20 / $-2.92$ / 1.7 &  4.59 (0.09, 33) &  4.58 (0.06,  7) &  2.34 (0.07, 10) &  2.21 (0.09, 23) \\
BD $-11^\circ$ 0145  & 4900 / 1.73 / $-2.18$ / 1.8 &  5.26 (0.09, 40) &  5.32 (0.04,  5) &  3.07 (0.07, 10) &  3.03 (0.05, 15) \\
CD $-24^\circ$ 1782  & 5140 / 2.62 / $-2.72$ / 1.2 &  4.81 (0.07, 43) &  4.78 (0.05,  6) &  2.64 (0.04,  9) &  2.54 (0.02,  8) \\
BS16550-087         & 4750 / 1.50 / $-3.33$ / 2.0 &  4.14 (0.09, 42) &  4.10 (0.14, 12) &  1.83 (0.04,  6) &  1.65 (0.13, 16) \\
\hline 
\end{longtable}
}

\end{appendix}

\end{document}